\documentclass[12pt]{iopart}
\usepackage{iopams}
\usepackage{epsfig}
 
\begin{document}
\tolerance = 10000
\title{Scaling of entanglement entropy in honeycomb lattice on a torus }

\author {Wen-Long You}
\address{School of Physical Science and Technology, Soochow
University, Suzhou, Jiangsu 215006, People's Republic of China}
\ead{wlyou@suda.edu.cn}
\date{\today}
\begin{abstract}
The entanglement entropy of noninteracting fermionic system confined to two-dimensional (2D) honeycomb lattice on a torus 
are calculated.
We find that the entanglement entropy can characterize Lifshitz phase transitions
without a local order parameter. In the noncritical phase and critical phase with nodal Fermi surface, the entanglement entropy satisfies the area law. The leading subarea term is a constant in gapped phase rather than a logarithmic additive term in gapless regime. The tuning of chemical potential allows for a nonzero Fermi surface, whose variation
along particular direction determines a logarithmic violation of the area law.
We perform the scaling of entanglement entropy numerically and find agreement between the analytic and numerical results. Furthermore, we clearly show that
entanglement spectrum is equivalent to edge spectrum.
\end{abstract}


\maketitle
\section{Introduction}
Entanglement entropy is a unifying theme in the interdisciplinary theories during the past decade. It is a helpful bridge
between quantum gravity, quantum information and
condensed matter physics. It is defined as
the von Neumann entropy of one of the reduced states
for a pure state. Given a normalized wave function $\vert \Psi_{AB} \rangle$ and a partition of the system into subsystems $A$ and $B$, the associated
reduced density matrix of $A$ part is obtained from
$\rho_A= \textrm{Tr}_B \vert \Psi_{AB} \rangle \langle\Psi_{AB} \vert$ by tracing out the degrees of freedom in $B$ part.
The bipartite entanglement entropy is given by,
\begin{eqnarray}
{\cal S}_{\rm vN}= -\textrm{Tr} \rho_A \log_2 \rho_A  .
\end{eqnarray}
Related generalized quantities are the R\'{e}nyi entropies defined as
\begin{eqnarray}
{\cal S}_{\rm n}= \frac{1}{1-n}  \log_2 \textrm{Tr}  \rho_A ^n,
\end{eqnarray}
whose limit recovers the von Neumann entropy via relation ${\cal S}_{\rm vN} = \lim_{n\to 1}{\cal S}_{\rm n}$. Once the complete values of ${\cal S}_{\rm n}$ for all $n$ are known, one can obtain all eigenvalues
of $\rho_A$, which are called the entanglement spectrum. The spooky nonlocal nature makes the entanglement entropy hold great advantage in the context of condensed matter physics. On one hand, it plays an essential role in determining the efficiency of numerical algorithms based on matrix product states, as well as an extension to tensor network states \cite{Vidal} and Projected Entangled Pair States (PEPS) \cite{Verstraete}. On the other hand, it is proven that the entanglement entropy is able to capture the quantum criticality in a many-body Hamiltonian \cite{Amico,Eisert}.
The constituents become extraordinarily entangled at criticality.
Quantum critical points are successfully pinpointed in
both free systems \cite{Osborne,Plenio} and interacting systems \cite{Wu,Barthel2,Haque,You1,You2},
especially for those critical ground states without prior knowledge of order parameters, e.g., disordered systems\cite{Refael,Laflorencie,Berkovits}
and quantum spin liquids \cite{Zhang}. In particular, the discovery of a topological phase transition, where distinct phases of matter have the same symmetry, was in urgent need of basis-independent way of identifying quantum critical points. In this respect, extensive work has utilized the phase sensitivity of the entanglement entropy
to successfully characterize the topological order.
The universal constant term in the
bipartite entanglement entropy has been
shown to be a probe of topological contribution \cite{Kitaev,Levin}.
The success of entanglement entropy as a diagnostic tool is essentially rooted in its peculiar scaling properties in a quantum state.
The growing awareness of the monotonic scaling of von Neumann entropy can be traced back to the study of black hole physics \cite{Bombelli,Srednicki,Hawking}, and then the profound relations to conformal field theory (CFT) in (1+1)-dimensions were established \cite{Holzhey,Calabrese2004}. Despite the formal simplicity, it is surprisingly difficult to calculate the entanglement entropy analytically, even for noninteracting models. The entanglement entropy in one-dimensional system has been extensively studied and the scalings have been explicitly established by CFT \cite{Amico,Vidal03}. Intensive research in terms of numerical algorithms unveiled a few qualitative and quantitative aspects in higher dimensions \cite{Kallin,Jiang}.

In this work, we study the half-space entanglement entropy of free fermion in the anisotropic honeycomb lattice with fine tuning chemical potential. In order to discriminate the relation between various boundaries and entanglement entropy, we place the fermion system on a torus. The scaling behavior of the entanglement entropy in a two-dimensional hexagonal lattice demands a comprehensive investigation.
The basic interest in honeycomb lattices comes from the graphene and associated Dirac fermion physics. 
More interestingly, a recent influx of ideas and toolboxes can create the so-called "artificial honeycomb lattice" \cite{Polini}, such as photonic lattices \cite{Rechtsman}, coupled molecular systems \cite{Gomes} as well as optical honeycomb lattice \cite{Tarruell}.
The manipulations of Dirac points are possible through
these artificial methods \cite{Lim,Montambaux1,Montambaux2}. The asymmetric tunneling amplitudes cause a topological Lifshitz transition from a semimetal to an insulator, which is characterized by
the topological change of the Fermi surface in the Brillouin zone \cite{Yamaji}. It was found that the Lifshitz points can be also captured by the entanglement entropy \cite{Rodney}. Moreover, based on the
bulk-edge correspondence \cite{Ryu2002,Hatsugai}, the entanglement entropy and the Berry phase are closely related in a topologically nontrivial system through the zero-energy edge states \cite{Ryu}. A fascinating aspect of graphene-like structure is that there are two types of edges, i.e., zigzag and armchair edges. The role of edges in determining the entanglement entropy in honeycomb lattice and associated scaling behavior are not fully understood. In addition, the chemical potential can be well tuned in graphene by a gate voltage, so the dependence of the entanglement entropy on the chemical potential is a nontrivial issue to be discussed.

 \section{Tight binding model on honeycomb lattice}
\begin{figure}[!ht]
\begin{center}
\includegraphics[width=8cm]{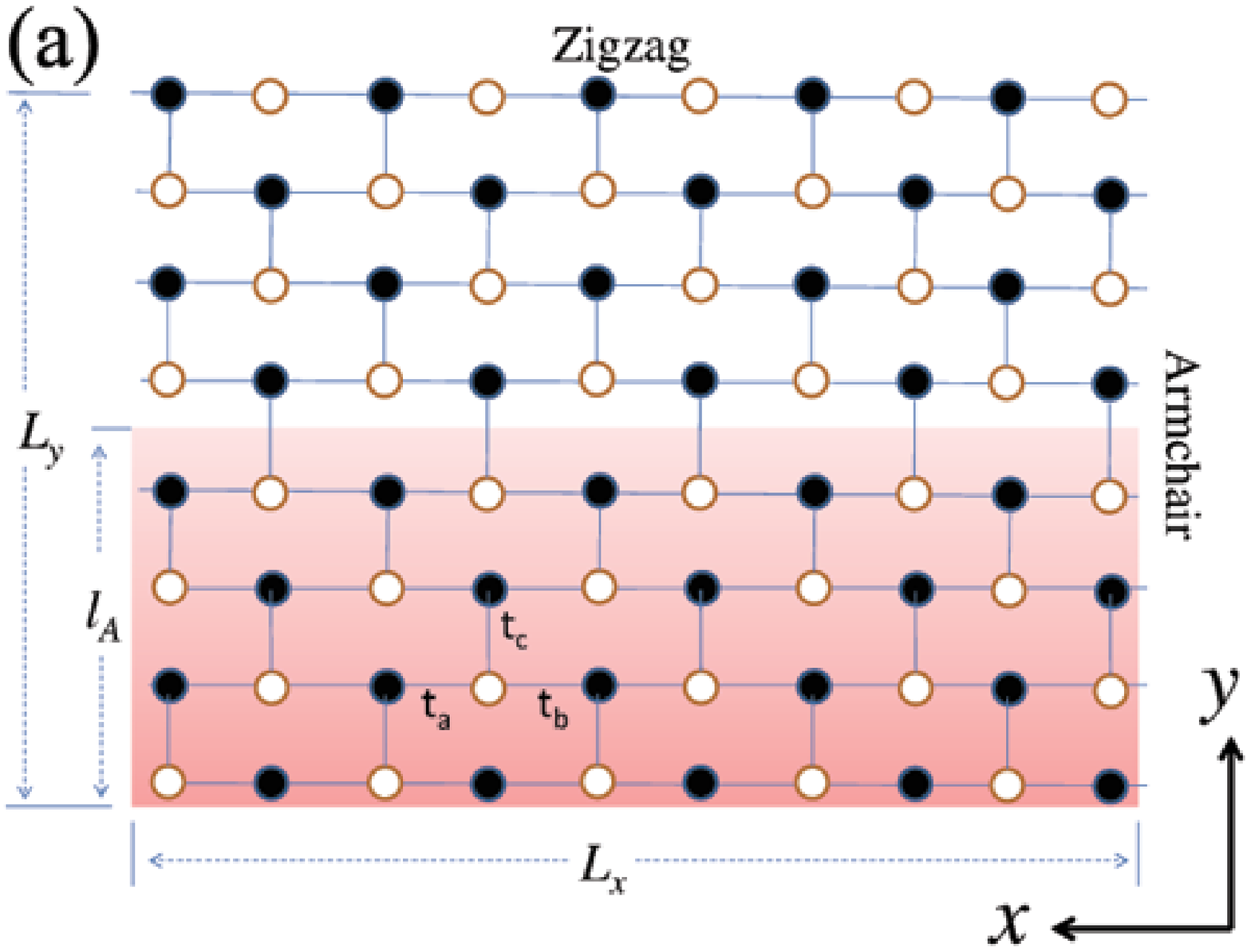}
\includegraphics[width=6cm]{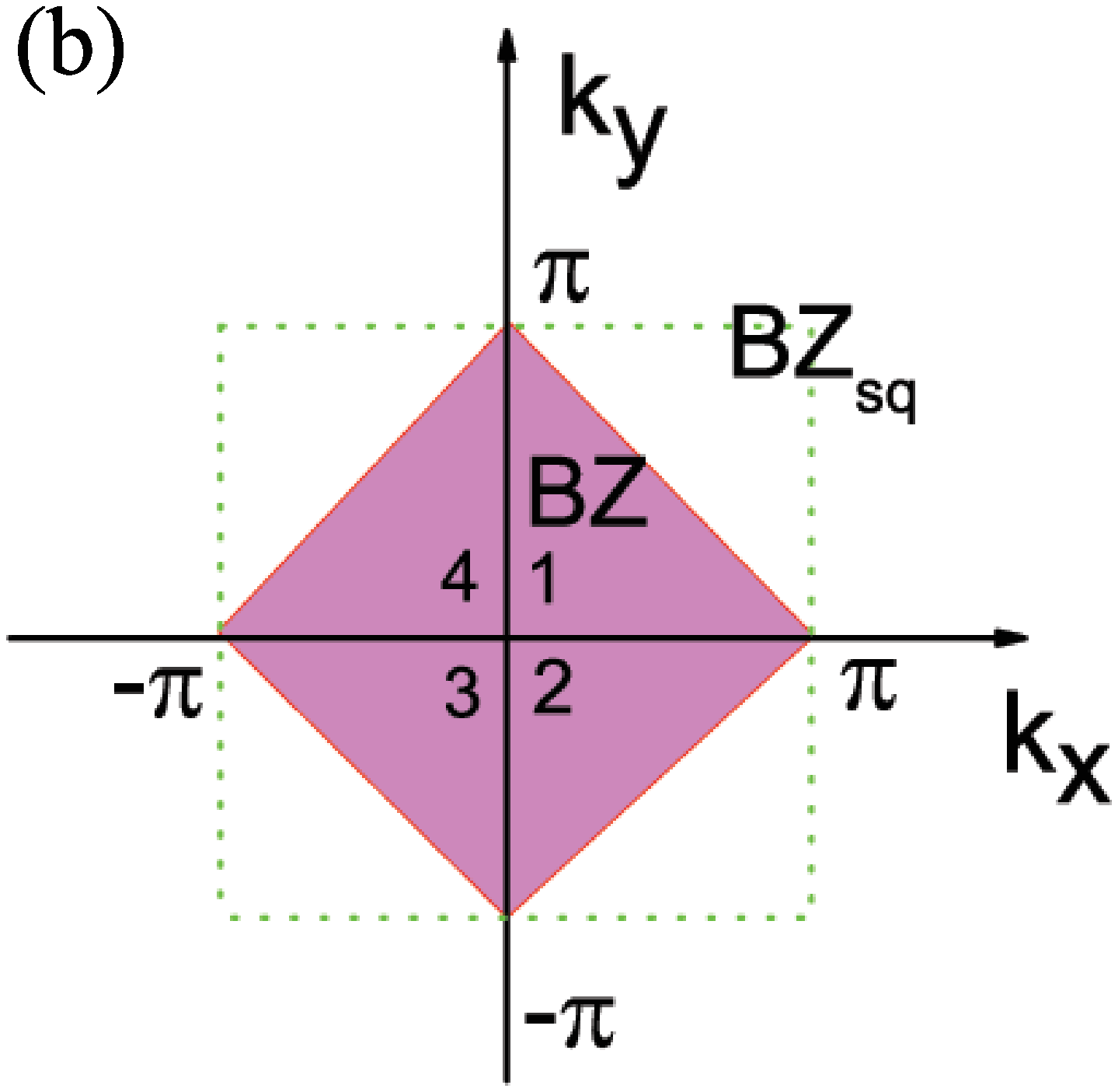}
\caption{(Color online) (a) Schematic illustration of brick-type
graphene torus. A virtual bipartition with zigzag edges is made horizontally.
The shaded region denotes one of the subsystems.
The dashed (red) box indicates a unit cell.
 A bipartition with armchair edges can be implemented by
 truncating the lattice vertically.
(b) $\mathbb{BZ}_{\rm sq}$ is the Brillouin zone of square lattice,
 and the double count for $\vec{k}$ and $\vec{k}+ \vec{\pi}$ indicates
 the $\mathbb{BZ}$ (region 1-4) is the Brillouin zone of the hexagonal lattice.
 }\label{schematicsofblockzigzag}
\end{center}
\end{figure}

\begin{figure}[t!]
\begin{center}
\includegraphics[width=7.5cm]{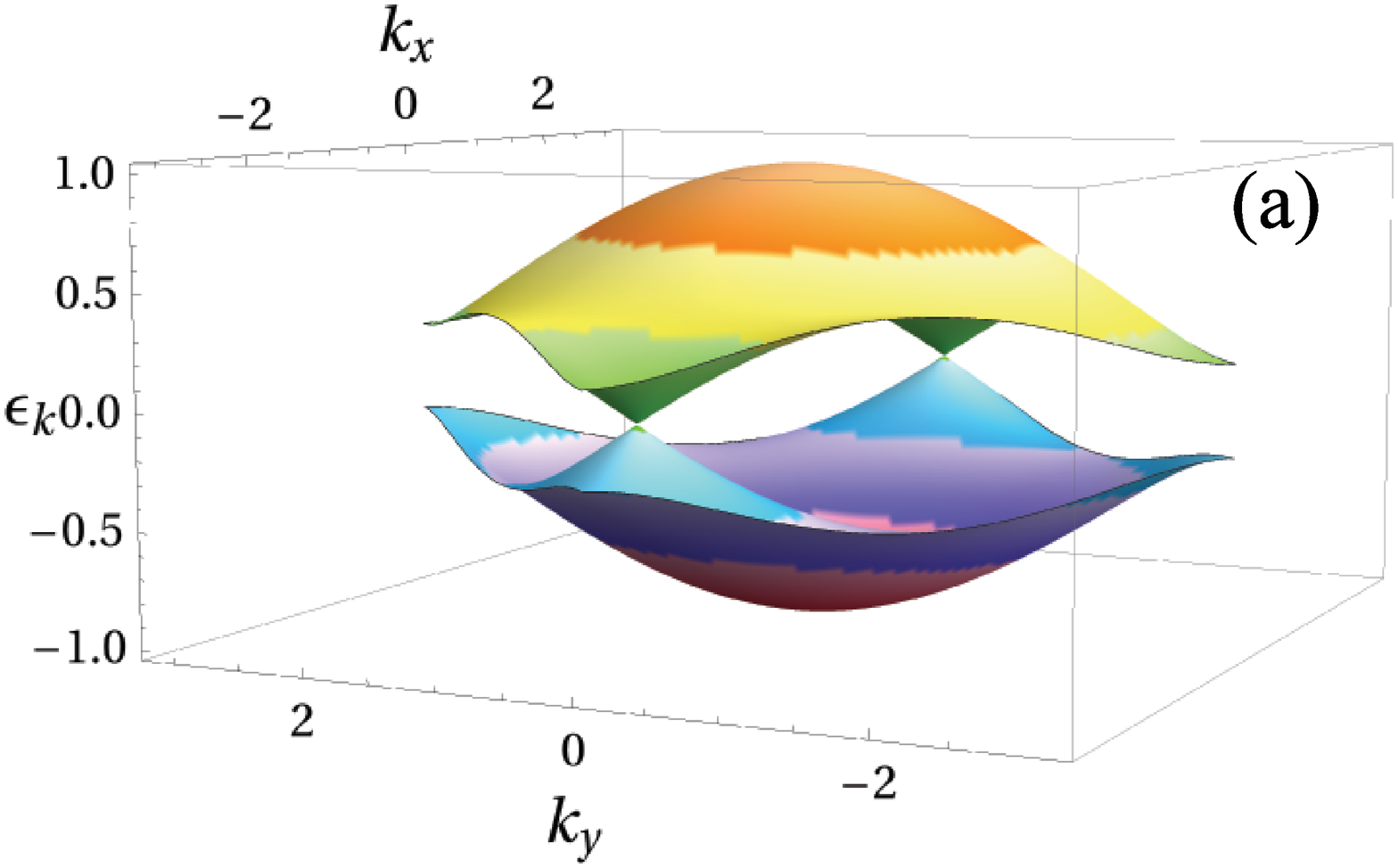}
\includegraphics[width=7.5cm]{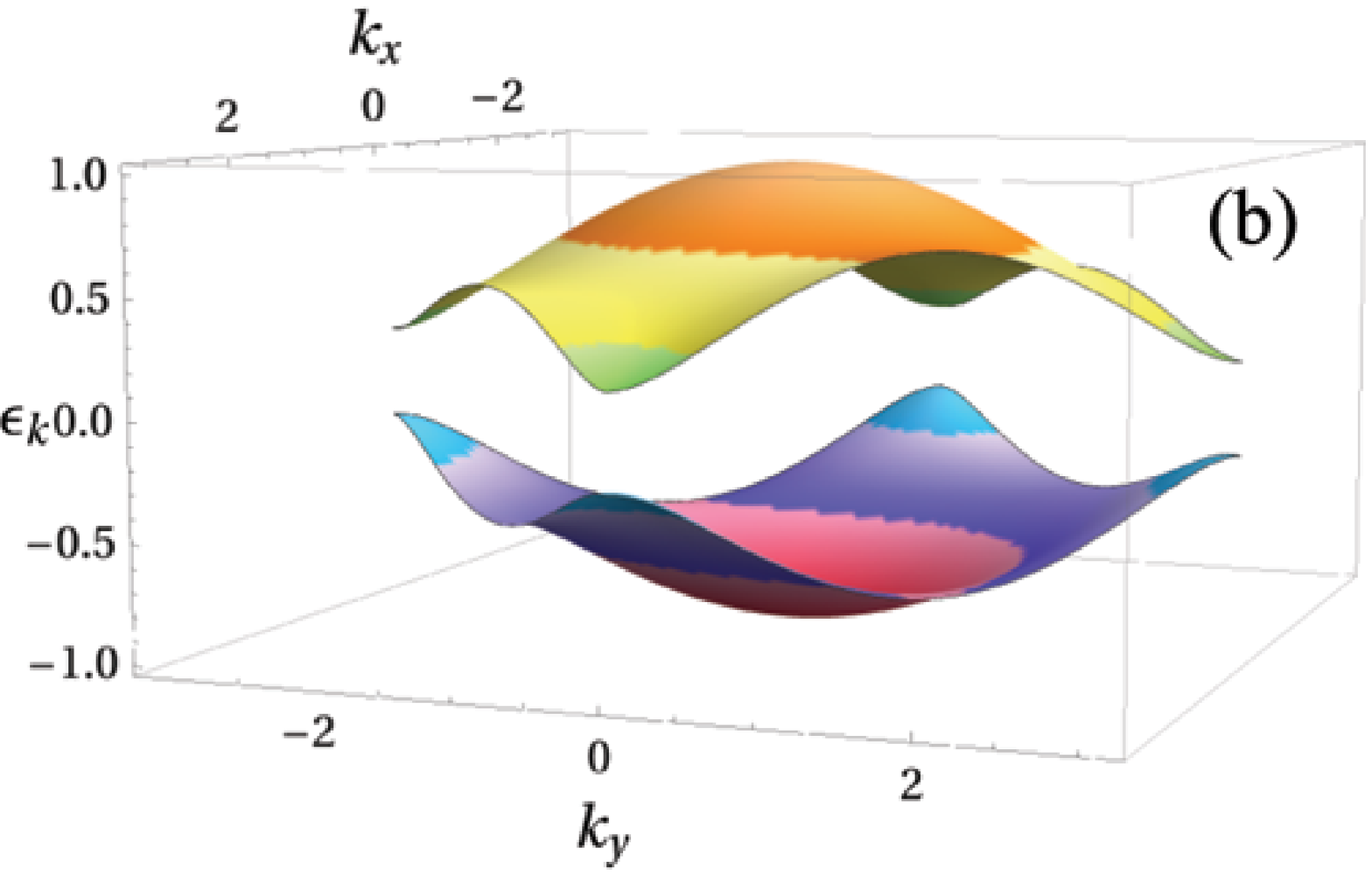}
\end{center}
\caption{(Color online) Three-dimensional view of the energy
spectrum of the Hamiltonian (\ref{bricktypegraphene}) for (a) $t_a=0.28$, $t_b=0.32$, $t_c=0.4$; (b) $t_a=0.15$, $t_b=0.25$, $t_c=0.6$.
}\label{EnergySpectrum}
\end{figure}

The anisotropic fermionic tight-binding model on a two-dimensional honeycomb lattice reads
\begin{eqnarray}
H=&-& \sum_{r\in \Lambda_A}\sum_{i=a,b,c} t_i (a_r^{\dagger} b_{r+\tau_i} +
b_{r+\tau_i}^{\dagger} a_r )-\mu \sum_{r\in \Lambda_A} a_r^{\dagger} a_{r}-\mu \sum_{r\in \Lambda_B} b_r^{\dagger} b_{r} ,   \label{bricktypegraphene}
\end{eqnarray}
where $a_r$ ($b_r$) is the annihilation operator on the triangular sublattice
 $\Lambda_A$ $\circ$ (respectively $\Lambda_B$ $\bullet$) of the $r$th dimer, $\tau_i$ is the nearest neighbor bond vector, and $\mu$ is the chemical potential. For simplicity, we confine the hexagonal lattice to the brick lattice without
changing lattice topology as shown in
Fig.\ref{schematicsofblockzigzag}(a). The nearest-neighbor separations
are given by
\begin{eqnarray}
\tau_a &=& (0, 1), \quad  \tau_b = (-1,0),
\quad \tau_c = (1, 0),\label{Bricksublatticeseparation}
\end{eqnarray}
where the bond length is set to unit. Taking Fourier transformation for all lattice sites, the Hamiltonian
(\ref{bricktypegraphene}) reduces to
\begin{eqnarray}
H&=& - \sum_{\vec{k}\in \mathbb{BZ}} {\bf c}_k^{\dagger} \left(\begin{array}{cc}
                                    \xi(\vec{k})-\mu & i \Delta(\vec{k}) \\
                                   - i \Delta(\vec{k}) & -\xi(\vec{k})-\mu
                                  \end{array}\right){\bf c}_k , \label{BdGHamiltonian}
\end{eqnarray}
and then the eigenspectrum is
\begin{eqnarray}
\epsilon(\vec{k})=\pm \sqrt{\xi^2(\vec{k})+\Delta^2(\vec{k})}-\mu,\label{epsilonk}
\end{eqnarray}
where $\xi(\vec{k})$= $t_a\cos k_x$ + $t_b \cos k_x$ + $t_c \cos k_y$,
$\Delta(\vec{k})$=$t_a \sin k_x$ - $t_b \sin k_x$ -$t_c \sin k_y$, ${\bf c}_{\vec{k}}^{\dagger}$=($c_{\vec{k}}^{\dagger}$, $c_{{\vec{k}}+{\vec{\pi}}}^{\dagger}$) and
$\vec{k}\equiv (k_x, k_y)$ is the Brillouin zone shown in
Fig.\ref{schematicsofblockzigzag}(b). The bulk band structure in Eq.(\ref{epsilonk})
is invariant under inversion :
$\vec{k} \to -\vec{k}$.
In the parameter space $\vert t_a-t_b \vert$ $<t_c<$ $t_a+t_b$
(containing symmetric case $t_a=t_b=t_c$), the valence and conduction bands intersect linearly at two $ \vec{k}_{\rm D}$-points (known as
the Dirac points), given by
\begin{eqnarray}
\vec{k}_{\rm D}=\frac{1}{2} \left( \begin{array}{c}
2 \pi-\arccos(\frac{t_c^2- t_a^2-t_b^2}{2 t_a t_b} ) \\
 \arccos( \frac{(t_a^2+t_b^2)t_c^2-(t_a^2-t_b^2)^2}{2 t_a t_b t_c^2})\\
                       \end{array}
                       \right),
\end{eqnarray}
which will be suited at $\vec{K}$-points, i.e., $(\pm \frac{2}{3}\pi,0$), of the first Brillouin zone for symmetric hoppings $t_1$ = $t_2$ = $t_3$.
The spectra and the corresponding Dirac points for a semimetallic phase is displayed in Fig.\ref{EnergySpectrum}(a).
The low-energy electronic states at any Dirac point are described by the massless relativistic Dirac fermions, which lead to a number of unconventional and fascinating phenomena, such as anomalous quantum Hall effect \cite{Ostrovsky}. The asymmetric hoppings make the two Dirac points move away from the $\vec{K}$-points and approach each other \cite{Wunsch}. At a critical asymmetry the Dirac points merge at one of the four inequivalent $\vec{M}$-points, that is, ($\pm\pi,0$) and ($0, \pm\pi$). A band gap opens upon increasing the asymmetry
further; see illustration in Fig.\ref{EnergySpectrum}(b). The topological invariant that characterizes the topological properties is given by the Zak phase, which is the integration of Berry connection over in
the reduced Brillouin zone \cite{Park,Delplace}. If the integral loop encloses
a gap closing point, the Zak phase gives
\begin{eqnarray}
\gamma &=&-i \int_{-\pi}^{\pi} \langle v_{-}(\vec{k}) \vert \partial_{\vec{k}} v_{-}(\vec{k})\rangle \cdot d\vec{k}=\pi,
\end{eqnarray}
where
$
v_{-}(\vec{k})=\frac{1}{\sqrt{2}}\left( \begin{array} {c}
1 \\
 e^{i \theta_k}\\
 \end{array} \right)
$
is the eigenvector of lower band and
$\theta_k  = \textrm{Arg}[\xi(\vec{k})+i\Delta(\vec{k})].$
Otherwise it is zero. The merging is a topological transition since the Berry phases $\pm \pi$ associated with the two Dirac points annihilate each other at the critical point.

\section{Entanglement entropy of noninteracting fermions}
The entanglement entropy in the ground state of free fermions can be determined from the correlation matrix as a fictitious Hamiltonian
defined as \cite{Peschel03,Peschel05,Peschel12,Chung01}
\begin{eqnarray}
\hat{G}_{ij}=\textrm{Tr}(\rho_A c_i^{\dagger} c_j),\label{GFofGraphene}
\end{eqnarray}
where $c_i$ ($c_i^{\dagger}$) is the annihilation (creation) operator for site $i, j \in A$. There exist an simple relation between correlation
matrix of order ${\cal O}(N_A)$ and the reduced density matrix of
order ${\cal O}(2^{N_A}) $ : $\rho_A$= $\det (1-\hat{G}) \exp\{ \sum_{ij}[\ln  \hat{G} (1-\hat{G})]_{ij}
c_i^\dagger c_j\}$ \cite{Cheong}, where $N_A$ is the number of sites in $A$ subsystem.
Calculating the eigenvalue $\lambda_n$ of matrix $\hat{G}$ requires
only the diagonalization of $ N_A \times N_A$ matrix, and then
the entanglement entropy is given by
\begin{eqnarray}
S_{\rm vN} =-\textrm{Tr}[\hat{G}  \log_2 \hat{G}  -(1-G) \log_2(1-\hat{G})]. \label{Svn-G}
\end{eqnarray}
The eigenvalue $\lambda_n$ ($0 \le$ $\lambda_n$ $\le$ 1) of
$\hat{G}$ is termed entanglement spectrum, which is an alternative description
of Li and Haldane's version \cite{Haldane}.  Recently the entanglement spectrum has been proposed as a ground-state property that captures characteristic edge excitations \cite{Fidkowski,Andreas,Thomale}.
In a large subsystem, most of the eigenvalues lie exponentially close to $0$ and  to $1$ \cite{Chung2011}. A special role is played by eigenvalues $\lambda_n =1/2$ corresponding to a zero-energy mode \cite{Turner}, and it will contribute a maximal value $1$ to the entanglement entropy by Eq. (\ref{Svn-G}).

In the following, we consider a
torus of size $L_x \times L_y$, and a subregion $A$
can be designated as $\Omega$= [0, $z_1 L_x$]$\times$[0, $z_2 L_y$]
with $z_i$ $\le$ 1. Here we measure the entanglement between a cornerless cylindrical region $A$
and its complement by two virtual cuts along $x$ or $y$ direction; see illustration in Fig.\ref{schematicsofblockzigzag}(a).
Zigzag edges emerge following a horizontal bipartition,
i.e., $z_1=1$, $z_2=1/2$, and $k_x$ labels the momentum running along the zigzag ribbon.
A vertical truncation produces armchair edges,
i.e., $z_1=1$, $z_2=1/2$, and $k_y$ is well defined instead.
One can simultaneously classify the entanglement spectrum with
respect to the momentum parallel to the cut,
and an intimate relation between entanglement spectrum
and eigenspectrum of subblock $A$ can be discovered.

\begin{figure}[!ht]
\begin{center}
\includegraphics[width=7.5cm]{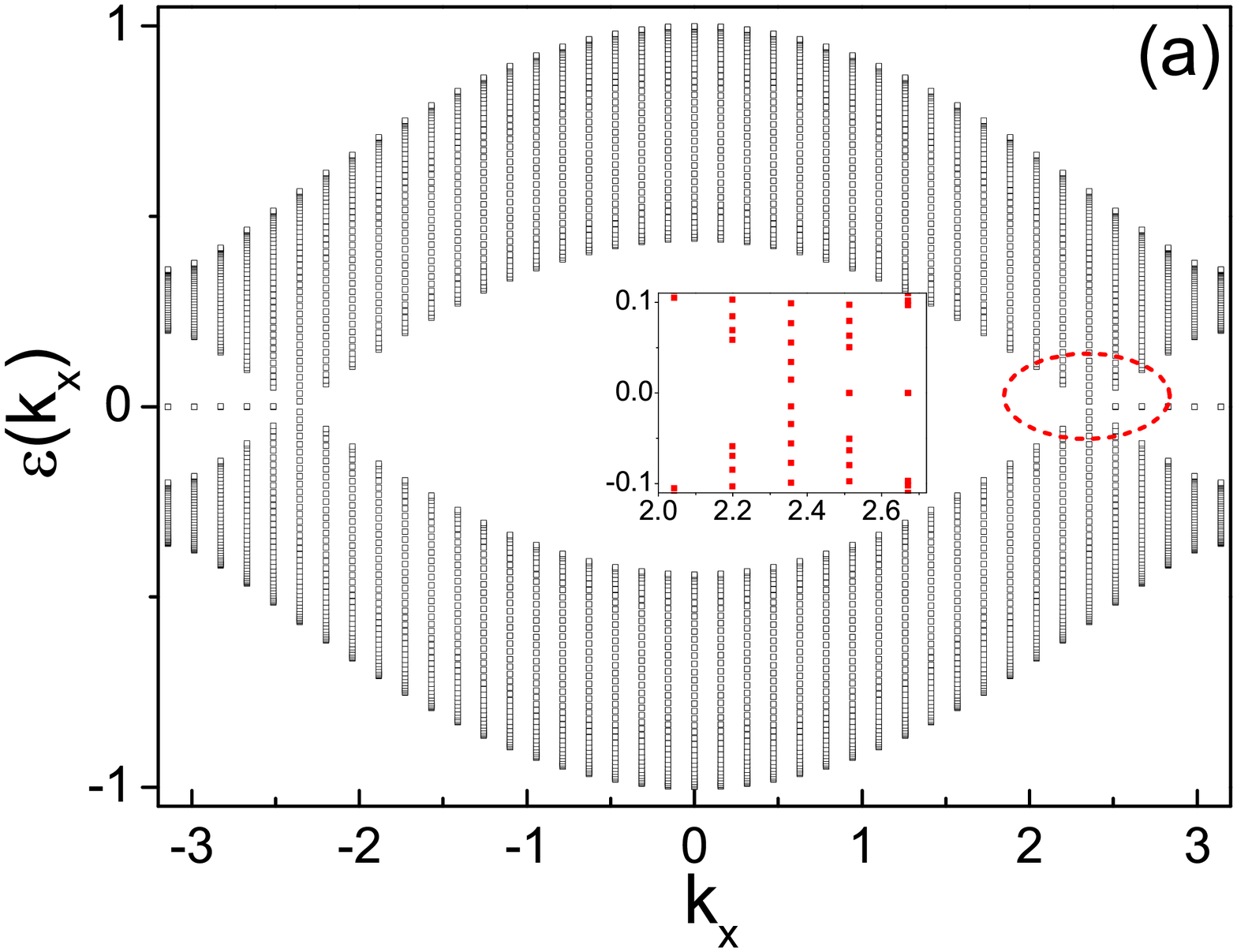}
\includegraphics[width=7.5cm]{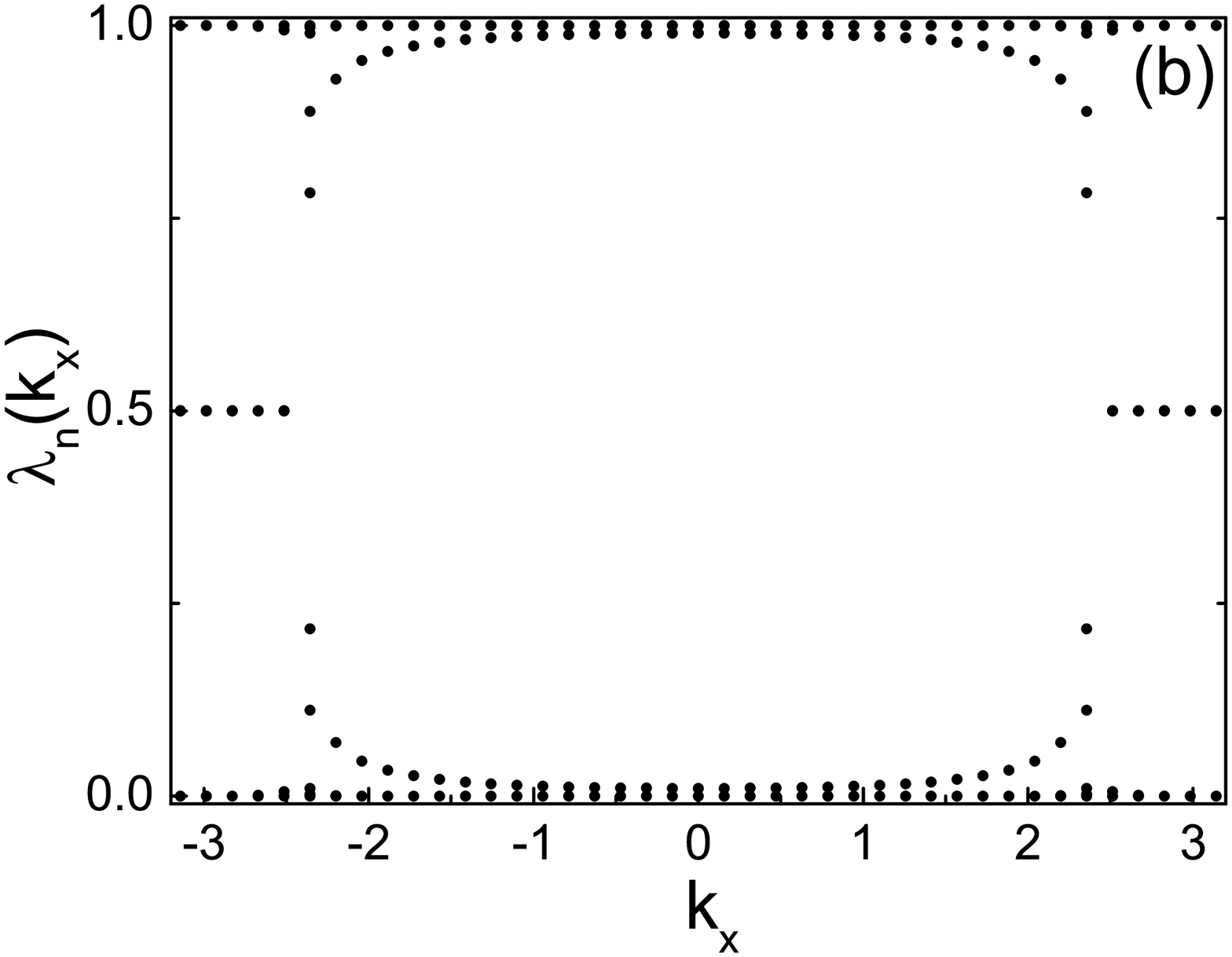}
\caption{(a) The energy spectrum for $t_a$=0.28,
$t_b$=0.32, $t_c$=0.40 on $L_x=40$, $L_y=80$ lattice with zigzag edges. (b) The corresponding entanglement spectrum at $\mu=0$. The inset amplifies the region in the dashed circle.}\label{ZigzagInMetal}
\end{center}
\end{figure}

\begin{figure}[!ht]
\begin{center}
\includegraphics[width=7.5cm]{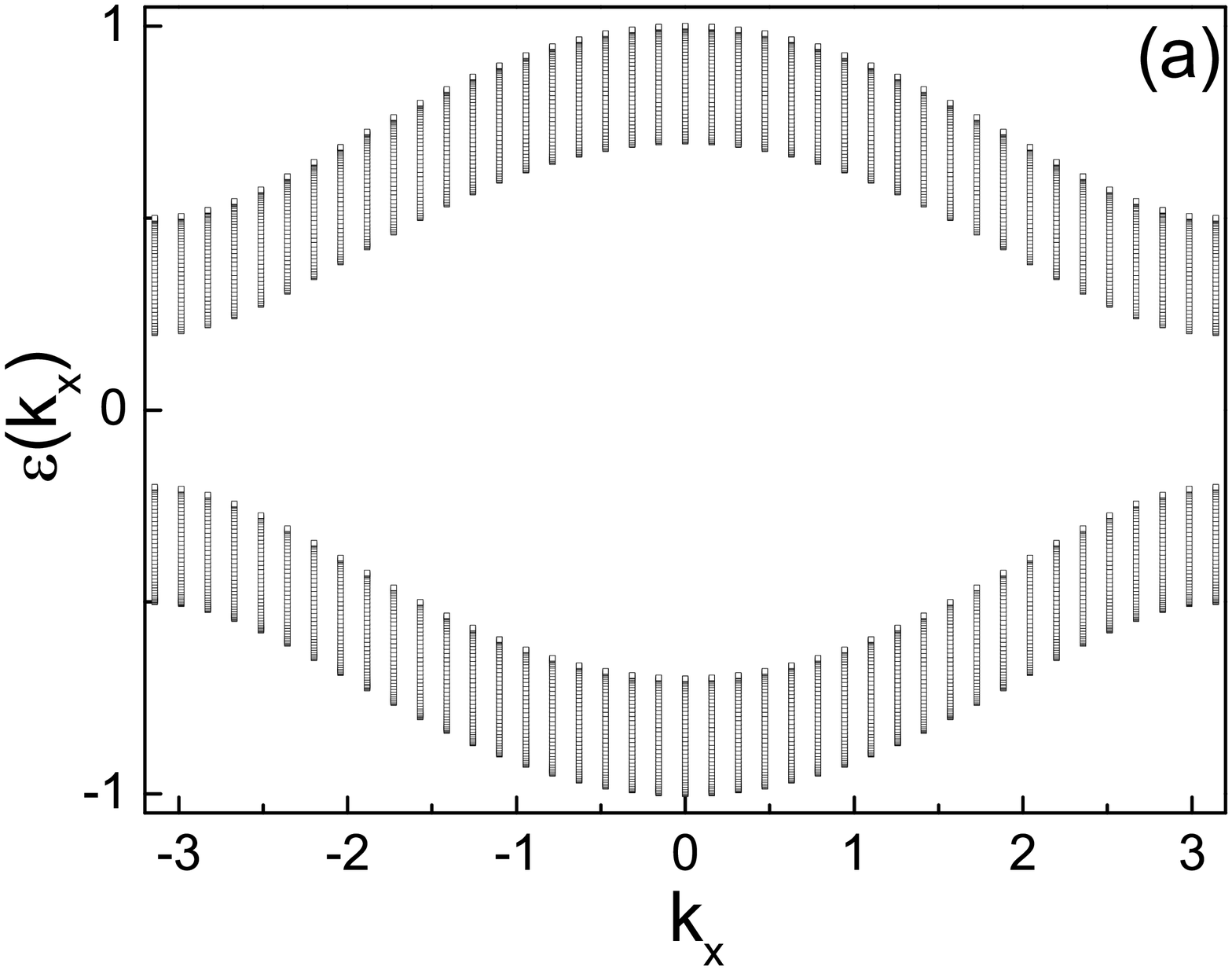}
\includegraphics[width=7.5cm]{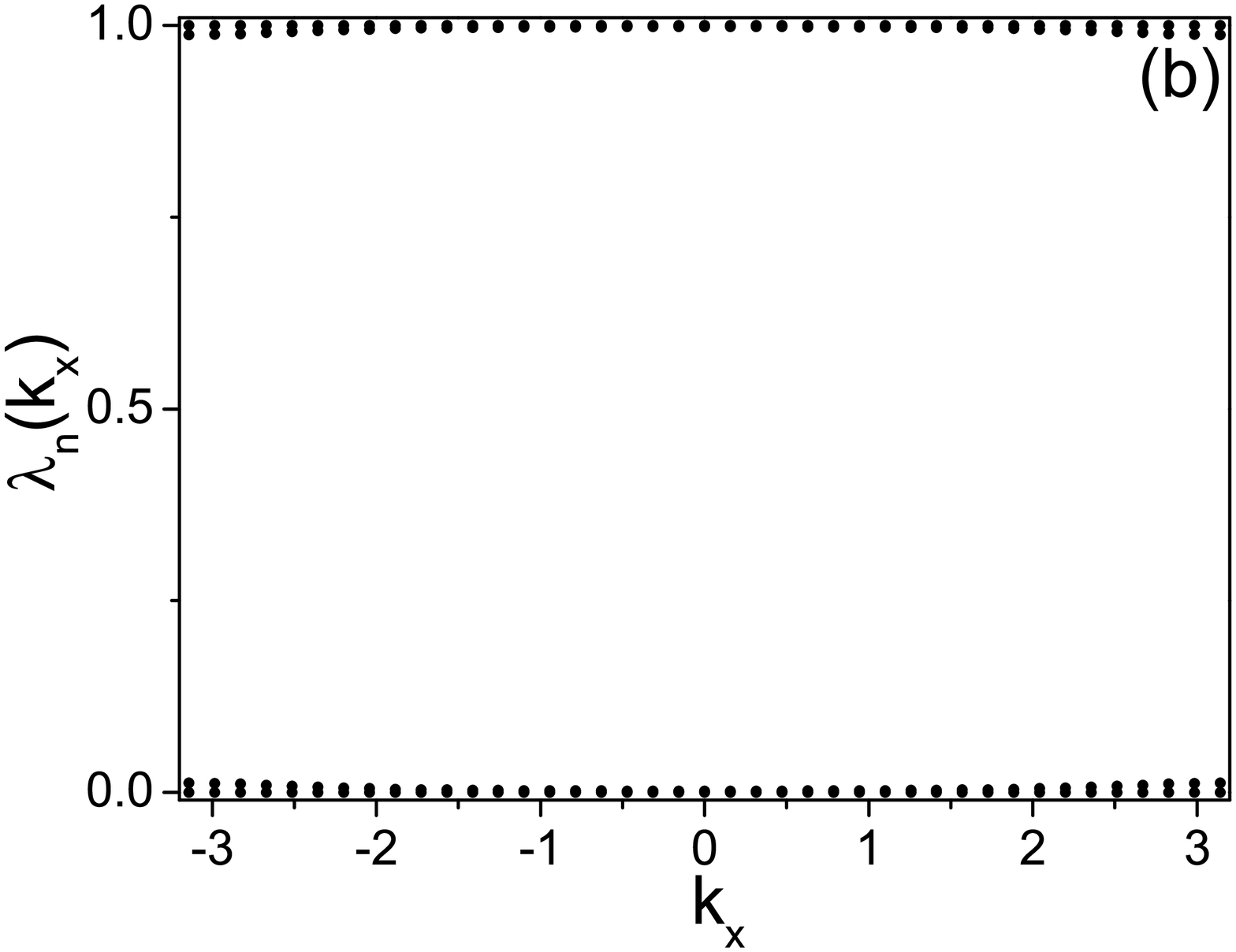}
\caption{(a) The energy spectrum for $t_a$=0.15,
$t_b$=0.25, $t_c$=0.6 on $L_x=40$, $L_y=80$ lattice with zigzag edges. (b) The corresponding entanglement spectrum at $\mu=0$.  }\label{ZigzagInInsulator}
\end{center}
\end{figure}

\begin{figure}[!ht]
\begin{center}
\includegraphics[width=7.5cm]{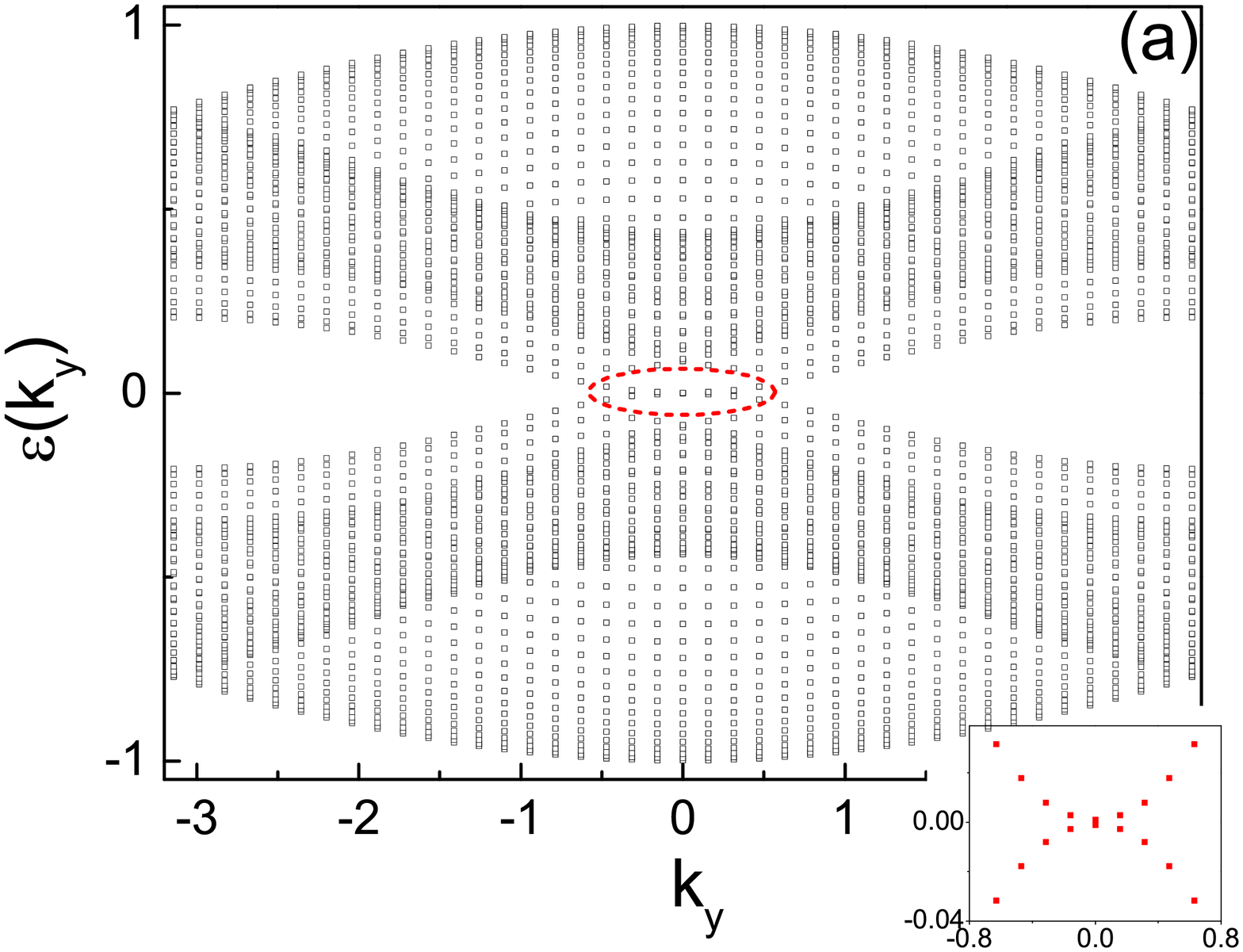}
\includegraphics[width=7.5cm]{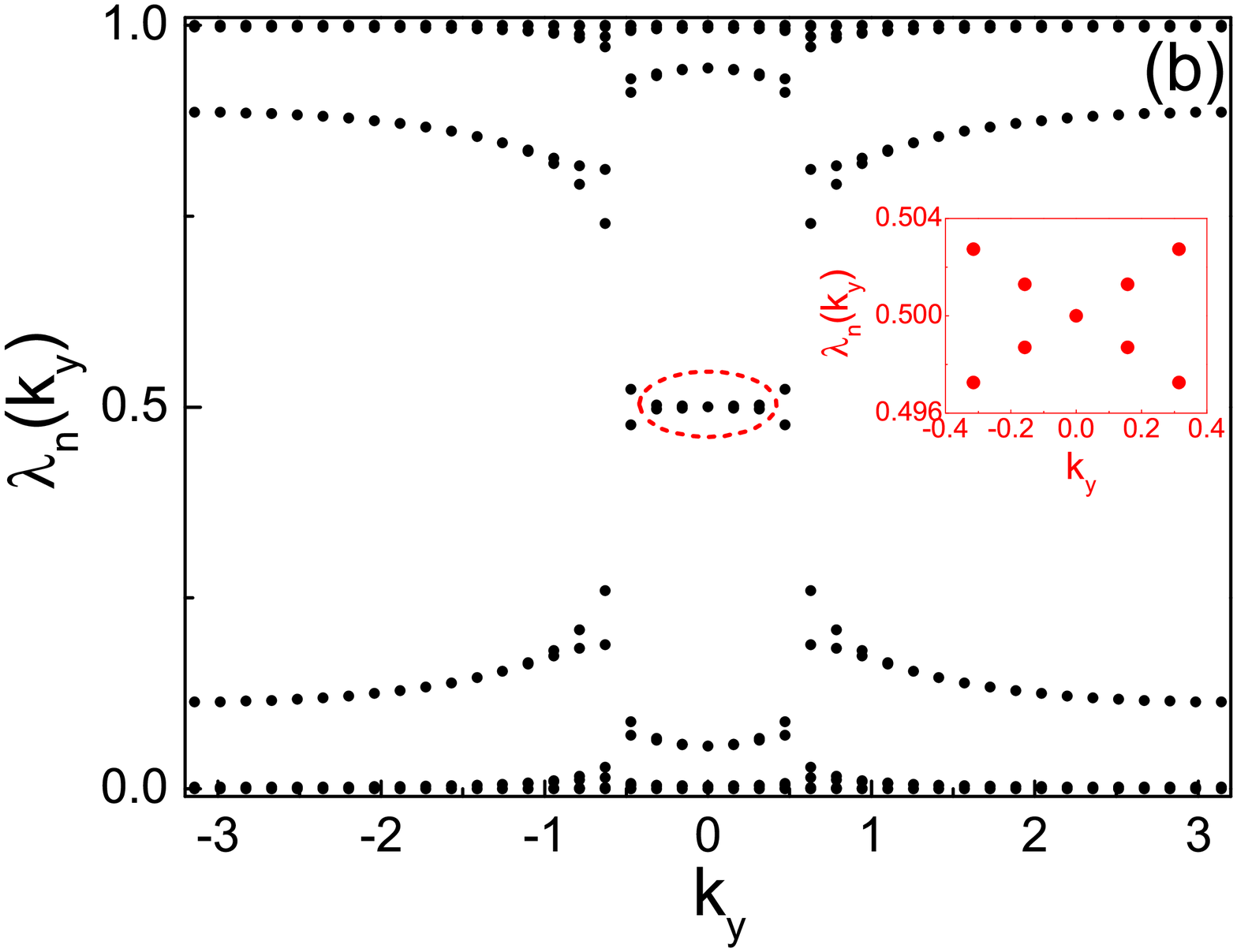}
\caption{ (Color online) The energy spectrum for $t_a$=0.28,
$t_b$=0.32, $t_c$=0.4 on $L_x=40$, $L_y=80$ lattice with armchair edges. (b) The corresponding entanglement spectrum at $\mu=0$. The inset amplifies the region in the dashed circle.}\label{ArmchairInMetal}
\end{center}
\end{figure}

\begin{figure}[!ht]
\begin{center}
\includegraphics[width=7.5cm]{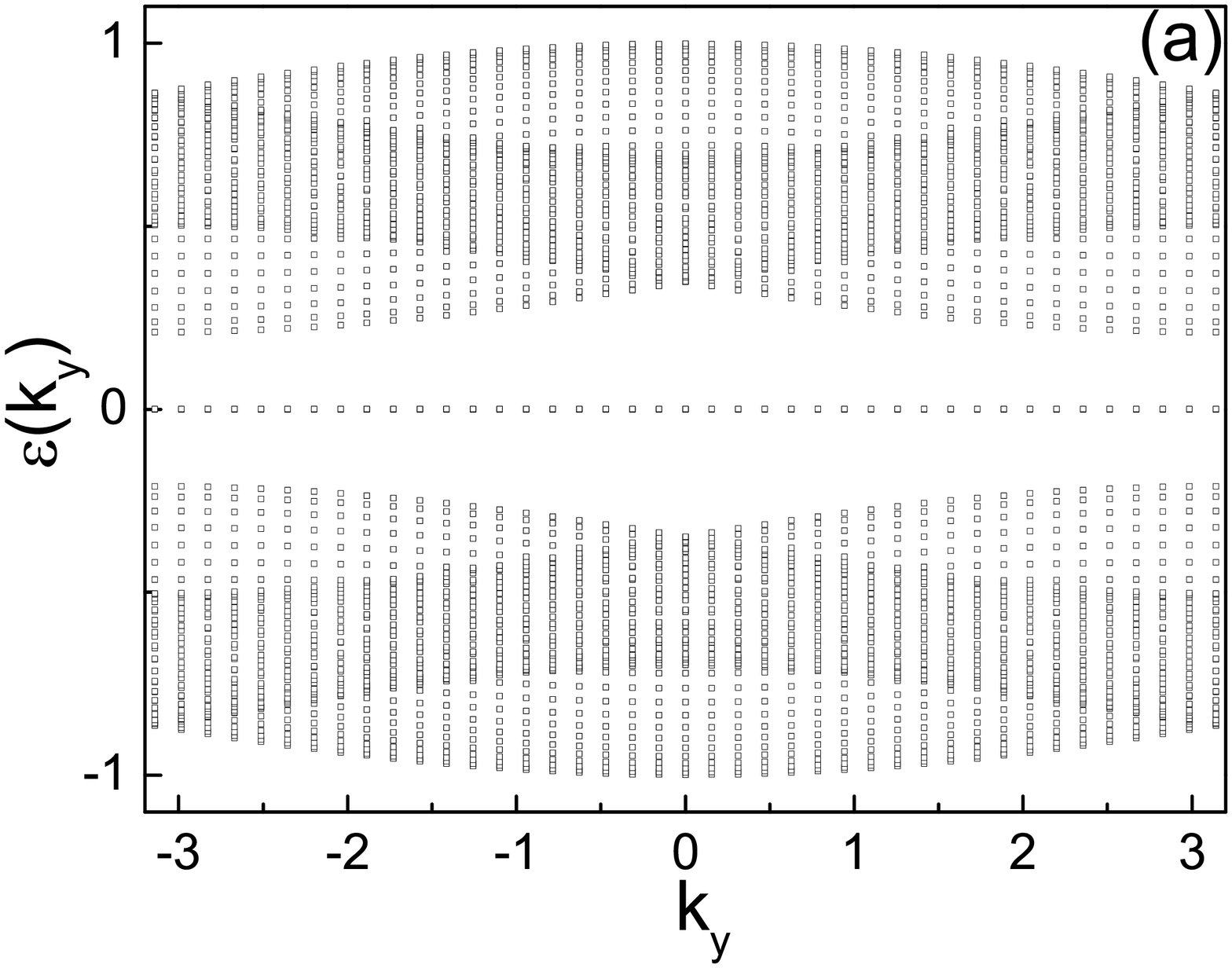}
\includegraphics[width=7.5cm]{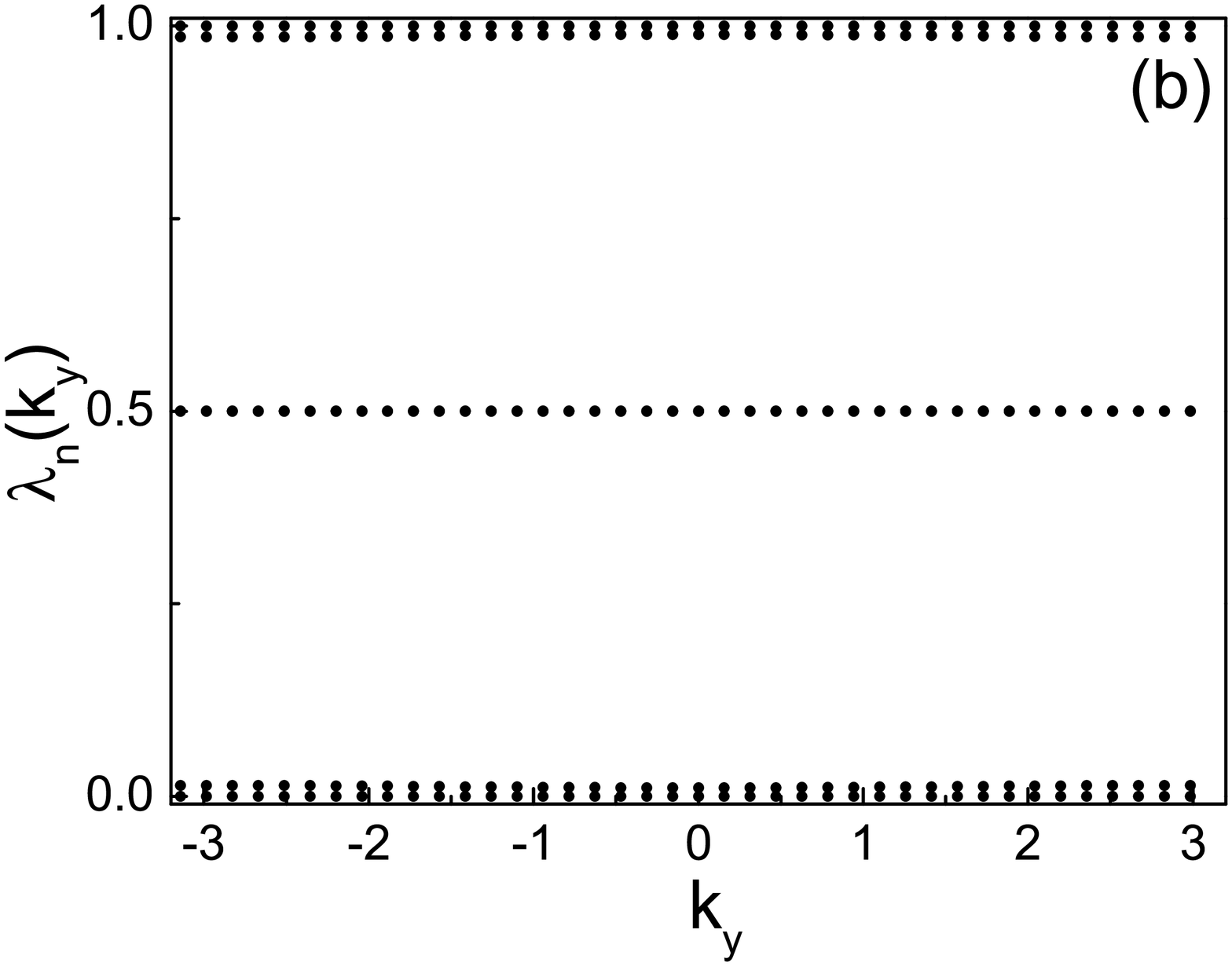}
\caption{ The energy spectrum for $t_a$=0.15,
$t_b$=0.25, $t_c$=0.6 on $L_x=40$, $L_y=80$ lattice with armchair edges. (b) The corresponding entanglement spectrum at $\mu=0$. }\label{ArmchairInInsulator}
\end{center}
\end{figure}

In Fig.\ref{ZigzagInMetal} (a), a flat band of
zero energy exists for momenta connecting with two Dirac points
along $x$ direction near the zigzag edges. Such zero-energy edge states are reflected by the maximally entangled
modes in entanglement spectrum, as shown in Fig.\ref{ZigzagInMetal} (b).
The fermion doubling in a chiral symmetric system is a universal property of the bulk, and the bulk-edge correspondence guarantees the edge states appear as pairs. In other words, such doubly degenerate entanglement modes are protected by
the chiral symmetry. In the semimetallic phase with zigzag edges, the contributions
to the entanglement entropy can be broadly separated into those from the edges and those from bulk, while only bulk states contribute to the entanglement
entropy in the insulating phase (see Fig.\ref{ZigzagInInsulator}).
However, the situation is quite different for armchair edges. The
interedge interaction of armchair edges in the semimetallic phase makes
the localized states fade away \cite{Arikawa}, and thus edge states
play a trivial role in the entanglement entropy (see Fig. \ref{ArmchairInMetal}).
Nevertheless, as is shown in
Fig.\ref{ArmchairInInsulator}, the anisotropy will induce number of edge states,
which contribute significantly to the entanglement entropy. It is explicit that the zero-energy states have one-to-one correspondences with the maximally entangled modes in the entanglement spectrum. In a sense, the entanglement spectrum is a very promising tool to characterize chiral edge states.

\begin{figure}[!ht]
\begin{center}
\includegraphics[width=9cm]{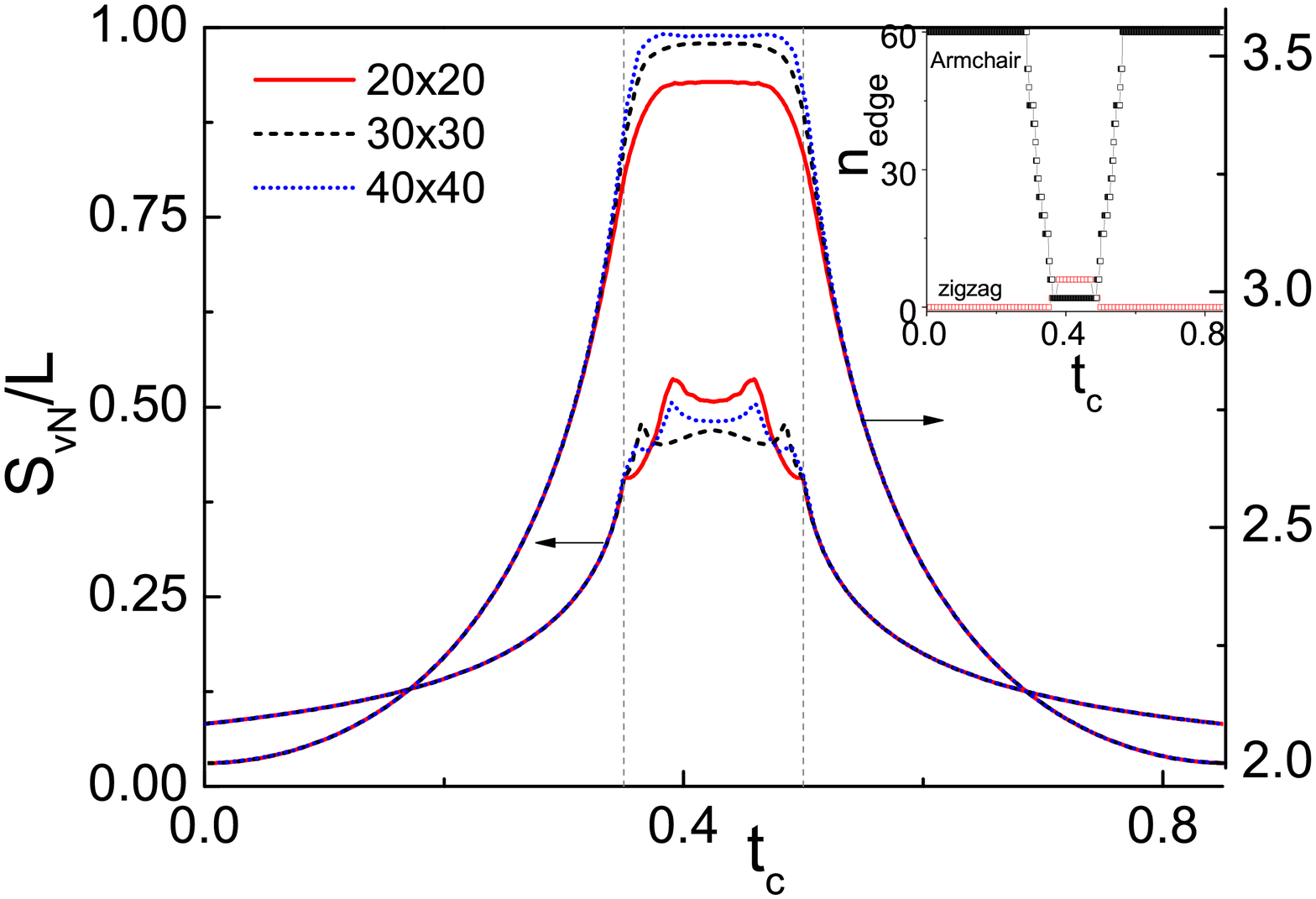}
\caption{(Color online) The entanglement entropy through path $t_a+t_b+t_c=1$ for  $t_a=0.15$,$\mu=0$ with zigzag edge (left axis) and armchair edge (right axis).  Inset shows the convolution of the number of edge states on $L_x=30$, $L_y=60$ lattice and $\mu=0$.
 }\label{EE-ta0.15tb0.25-tc}
\end{center}
\end{figure}

We show the half-space entanglement entropy as a function of $t_c$ along the normalized path $t_a+t_b+t_c=1$ for $t_a=0.15$, $\mu=0$ in Fig.\ref{EE-ta0.15tb0.25-tc}. The entanglement entropy changes discontinuously when $t_c$ moves across points $t_{c}^{1}=\vert t_a -t_b \vert$ and $t_{c}^{2}=t_a+t_b$.
In fact, the abrupt change associated with the topological phase transition can be grasped more clearly by its first derivative for both types of bipartition and any $\mu$ within the bulk bands. The system is an insulator when $t_{c}<\vert t_a -t_b\vert $ or $t_{c}>$$t_a +t_b$, and enters the semimetallic phase when $\vert t_a -t_b\vert$$<t_{c}<$$ t_a +t_b$. In the insulating phase, the entanglement entropy is linear with the block length for both edges, and the deviation seems negligible. However, the finite-size effect is prominent in the gapless phase.
There exists a remarkable correction to the linear dependence on the block length. We note that the correction is likely a logarithmic form for armchair edges, whereas an oscillating correction exists for zigzag edges. We attribute this peculiar correction to finite-size effect of edge modes. The trends in the proliferation of edge modes for both edges are converse as we approach a quantum critical point, as displayed in the inset of Fig.\ref{EE-ta0.15tb0.25-tc}. We find that the existence of edges state is that $t_a$, $t_b$ and $t_c$ should satisfy the triangle inequality for zigzag edge \cite{Kohmoto}.

Next we investigate the scaling behavior of bipartite entanglement entropy in the presence of different edges. The scaling of the entanglement entropy has attracted much attention by a flurry of recent works \cite{Barthel,Klich09,Song11,Song12}. A general form of the entanglement entropy in two-dimensional system is argued to be of the form
\begin{eqnarray}
S_{\rm vN}&=& c_1 L \log_2 L + c_2 L + c_3 \log_2 L + c_4, \label{Scalingin2D} \label{logarithmicscaling}
\end{eqnarray}
where $c_i$ ($i$=1,$\cdots$, 4) are size independent coefficients. Eq.(\ref{Scalingin2D}) includes a universal logarithmically divergent correction and a nonuniversal area-law term. Which term is the leading term requires further elucidation of the Fermi surface. We will give both analytical and numerical results for the logarithmical scaling of entanglement entropy in one concrete case.

\begin{figure}[!ht]
\begin{center}
 \includegraphics[width=7.5cm]{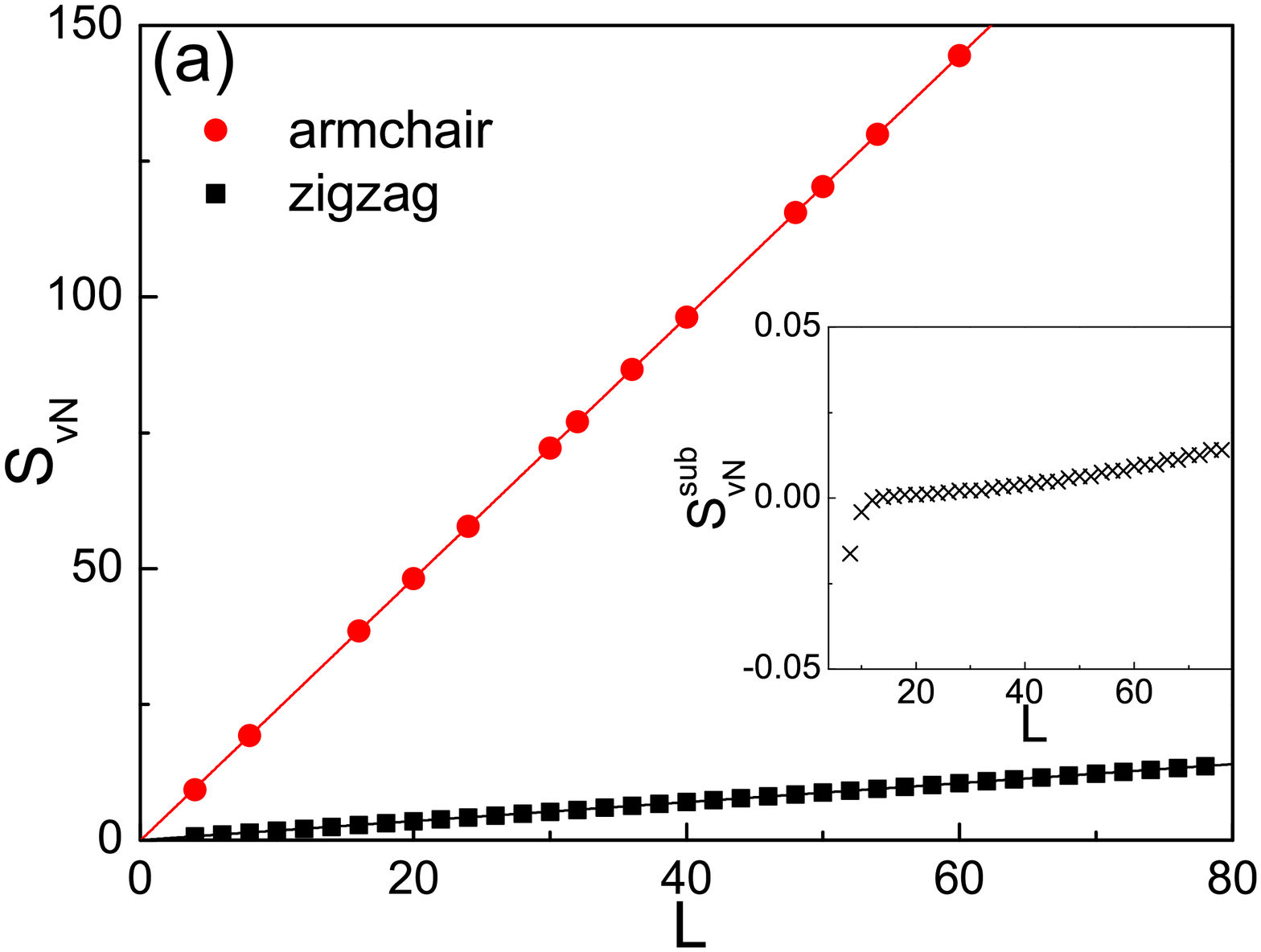}
\includegraphics[width=7.5cm]{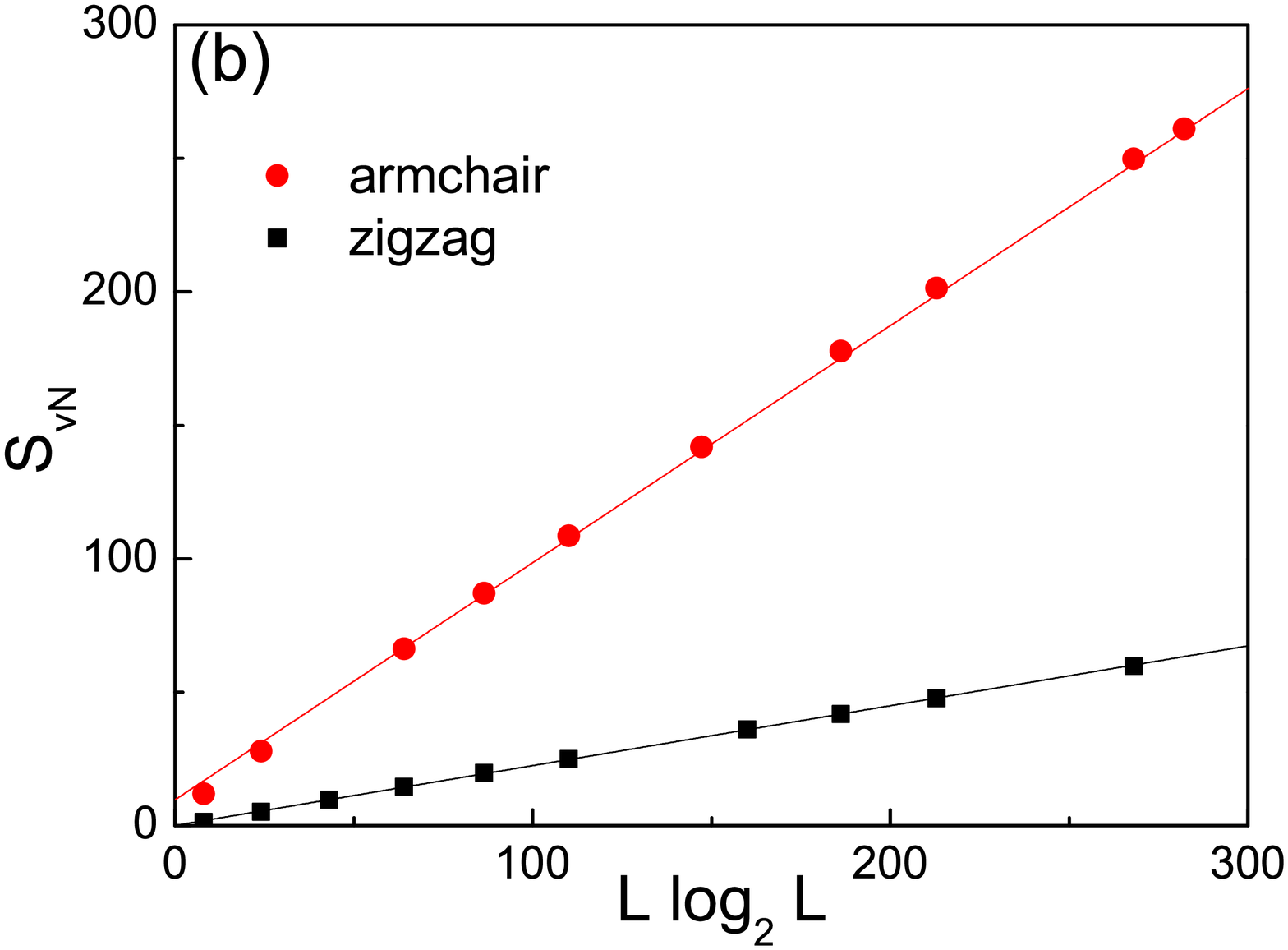}
\caption{(Color online) The scaling of entanglement entropy at $t_a$=0.15,
$t_b$=0.25, $t_c$=0.60 when (a) $\mu=0$ and (b) $\mu=-0.5$. Lines are linear fits. $S_{\rm vN}^{\rm sub}$ for $t_a$=0.15,
$t_b$=0.25, $t_c$=0.60 with zigzag edge is shown in the inset. }\label{EE-ta0.15tb0.24tc0.6}
\end{center}
\end{figure}

\begin{figure}[!ht]
\begin{center}
 \includegraphics[width=7.5cm]{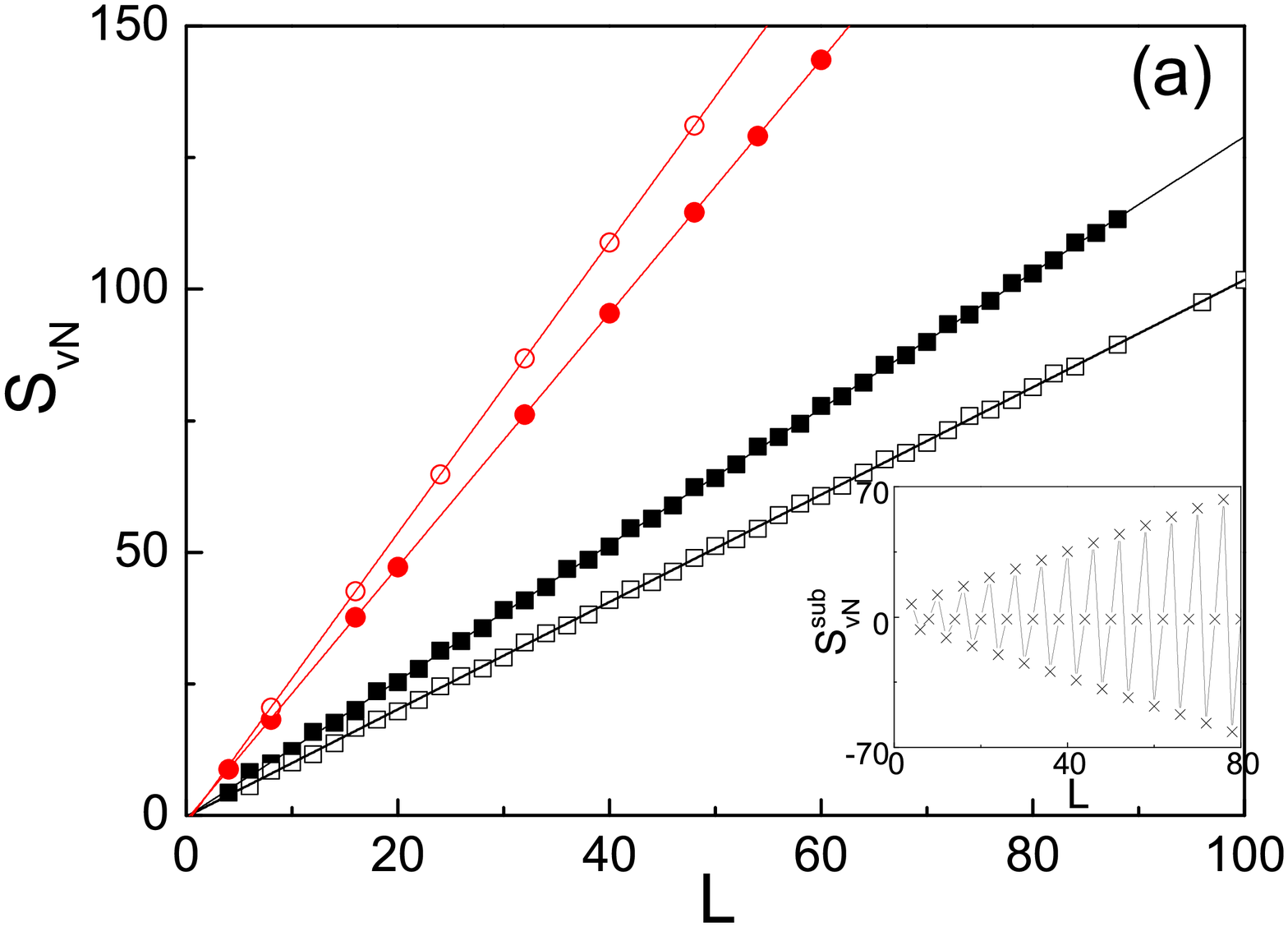}
\includegraphics[width=7.5cm]{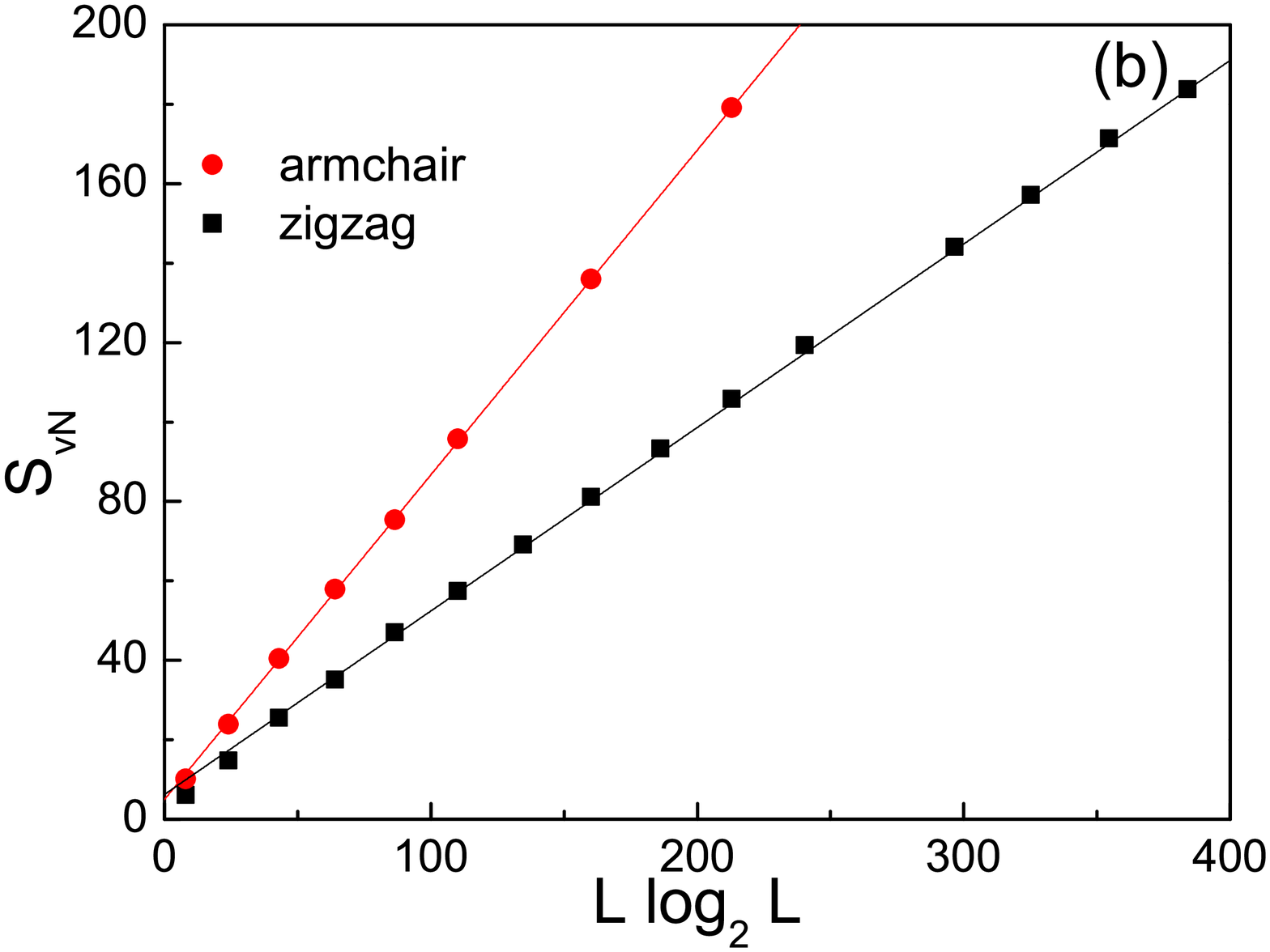}
\caption{(Color online)(a) The scaling of entanglement entropy with
zigzag (square) and armchair edges (circle) at $t_a$=$t_b$=$t_c$=1.0 (solid) and $t_a$=0.28, $t_b$=0.32, $t_c$=0.40 (open) with $\mu=0$. $S_{\rm vN}^{\rm sub}$ for $t_a$=$t_b$=$t_c$=1 with zigzag edge is shown in the inset. (b) The scaling of entanglement entropy with
both zigzag and armchair edges at $t_a$=$t_b$=$t_c$=1.0 with $\mu=-1.0$. Lines are linear fits. }\label{EE-ta1tb1tc1}
\end{center}
\end{figure}

Apparently in the insulating phase, the entanglement entropy
grows asymptotically as $L$ \cite{Wolf1}, as supported in Fig.\ref{EE-ta0.15tb0.24tc0.6}(a).
For $\mu=0$ in the semimetallic phase, the Fermi surface reduces to zero dimension, and then the entanglement converges to the area law \cite{Barthel,Levine,Ding}, as is shown in
Fig.\ref{EE-ta1tb1tc1}(a). We see that the slope of scaling changes for varying  parameters in the semimetallic phase, so the area-law term is nonuniversal. Apart from the leading power-law term, the subleading term is of general interest. To this end, we plot the quantity
\begin{eqnarray}
S_{\rm vN}^{\rm sub}= L S_{\rm vN}(L+2)-(L+2)S_{\rm vN}(L),
\end{eqnarray}
which will cancel the area-law term and hence extract the subleading contribution. It is remarkable that $S_{\rm vN}^{\rm sub}$ behave diversely in different phases. If the entanglement entropy follows a behavior as $
S_{\rm vN}(L)= c_2 L + c_4 + O(1/L)$,
$S_{\rm vN}^{\rm sub}$ will approach a 
constant term like
\begin{eqnarray}
S_{\rm vN}^{\rm sub} \sim -c_4,
\end{eqnarray}
as is shown in the inset of Fig.\ref{EE-ta0.15tb0.24tc0.6}(a). In contrast, $S_{\rm vN}$ displays a divergent oscillation in the inset of Fig.\ref{EE-ta1tb1tc1} (a). A careful analysis identifies a logarithmic additive term in these
critical phases by following $S_{\rm vN}(L)= c_2 L + c_3 \log_2 L + O(1)$,
and then the subarea law of the entanglement entropy yields\cite{Ding,Rong}
\begin{eqnarray}
S_{\rm vN}^{\rm sub} \sim -c_3 \log_2 L.
\end{eqnarray}
The subdominant contribution stems from one chiral mode in the zero-dimensional Fermi
surface, which infers $c_3$ = 1/3 by the Fisher¨CHartwig conjecture \cite{Calabrese12,Calabrese11}.
The oscillation can be interpreted as the following: In case of symmetric hoppings, the existence of zero-energy localized states is restricted for $1/3$ of the total momentum parallel to the zigzag edge, and hence a multiple of $3$ in block length favors more edge states. Hence, the oscillation has a period of $3$, as is observed in the inset of Fig.\ref{EE-ta0.15tb0.25-tc}.

A significant difference of the scaling in the entanglement entropy is found in the presence of finite density of state at the Fermi surface. In such situation, every low-energy radial mode is approximately treated as a chiral excitation that contribute a ln$L$ to the total entanglement entropy \cite{Swingle}. Therefore, a multiplicative logarithmic correction to the power-law behavior arises for finite Fermi surface (i.e., line nodes), which are reported in Fig.\ref{EE-ta0.15tb0.24tc0.6}(b) and Fig.\ref{EE-ta1tb1tc1}(b)\cite{Wolf2,Weifei}. Noticeably, we see that the entanglement entropy with armchair edge is larger than that with zigzag edge, and the reason will be accounted soon.

Widom conjecture \cite{Widom} predicted that the prefactor of the logarithmic term is determined by chemical potential $\mu$ via an accurate formula \cite{Gioev}
\begin{eqnarray}
c_1(\mu)&=& \frac{1}{24 \pi} \int_{\partial \Omega} d S_x\int_{\partial \Gamma(\mu)}dS_k \vert   \mathbf{n_x} \cdot  \mathbf{n_k} \vert,
\label{c1mu}
\end{eqnarray}
where $\Omega$ is the real-space region of $A$, rescaled by $L$ such
that Vol($\Omega$)=1. Vectors $\mathbf{n_x}$ and $\mathbf{n_k}$ denote
the normal vectors on the real-space surface $\partial \Omega$ and the
Fermi surface $\partial \Gamma(\mu)$.
From Eq.(\ref{c1mu}), we see the topology of the subsystem and the Fermi surface
has intimately related to the entanglement entropy. Thus the existence of edge states
crucially affects the physical properties of graphene
system. Besides, it was argued that Lifshitz quantum phase transitions are accompanied with a nonanalyticity in
the prefactor of the leading-order term \cite{Cramer}.
In two dimensions the calculation of $c_1(\mu)$ can be simplified. The normal unit vectors in real space are $\pm \hat{y}$ for zigzag-type bipartition
and $\pm \hat{x}$ for armchair-type bipartition.
Since the Fermi surface is a curve in two dimensions, we can
parameterize the curve by $k_x = k_x(\theta)$ and $k_y = k_y(\theta)$ like \cite{Rodney}
\begin{eqnarray}
(t_a +t_b) \cos k_x   +  t_c \cos k_y &\equiv& \mu \cos \theta ,\nonumber\\
(t_a -  t_b) \sin k_x  - t_c \sin k_y &\equiv& \mu \sin \theta . \nonumber
\end{eqnarray}
Then we have
\begin{eqnarray}
c_{1,\textrm{zigzag}}(\mu)&=&\frac{1}{12\pi}\int_{\partial \Omega}  d\theta  \left\vert\frac{dk_x}{d\theta}\right\vert , \label{zigzagcmu}\\
c_{1,\textrm{armchair}}(\mu)&=&\frac{1}{12\pi}\int_{\partial \Omega}  d\theta  \left\vert\frac{dk_y}{d\theta}\right\vert \label{armchaircmu} .
\end{eqnarray}
Eq.(\ref{zigzagcmu}) and Eq.(\ref{armchaircmu}) indicates $c_{1}(\mu)$ are proportional to the variation of Fermi surface along $x$ and $y$, respectively.
For $\vert \mu \vert $ is large, the Fermi surface is an oval,
while the disconnected Fermi components appear
with the decrease of $\vert \mu \vert$. Without loss of generality, we give a simple demonstration of the solution for symmetric case $t_a=t_b=t_c=1$. The dispersion has particle-hole symmetry and 90-degree rotational invariance, so we only consider positive $\mu$ in the first quadrant of the Brillouin zone,
and then the Fermi surface can be parameterized by
\begin{eqnarray}
\left\{
\begin{array}{l} \nonumber
k_x(\theta)=  \cos^{-1}  R(\mu,\theta)     \\
k_y(\theta)=  \sin^{-1} (- \mu \sin \theta )     \\
\end{array} \right.
\end{eqnarray}
for $0 \le \mu \le 1$ and $\theta \in (-\pi,0)$. Here $R(\mu,\theta)
=(-\sqrt{1-\mu^2 \sin^2 \theta}+\mu \cos \theta)/2 $. The longitudinal variation $\Delta k_y$ is $2 \arcsin(\mu)$ and the
horizontal variation $\Delta k_x$ is $\arccos(\frac{-\mu-1}{2})-\arccos(\frac{\mu-1}{2})$. When
$\mu>\sqrt{5}$, it is replaced by
\begin{eqnarray} \nonumber
\left\{
\begin{array}{l}
k_x(\theta)=  \cos^{-1} R(\mu,\theta)     \\
k_y(\theta)=     \sin^{-1} (- \mu \sin \theta )   \\
\end{array} \right.
\end{eqnarray}
for $\theta \in [-\arccos (\frac{\mu^2+3}{4 \mu}),0]$. The longitudinal variation $\Delta k_y$ is 
$\arccos \frac{\mu^2-5}{4} $ and the
horizontal variation $\Delta k_x$ is $\arccos(\frac{\mu-1}{2})$.
Otherwise when $1<\mu<\sqrt{5}$,
\begin{eqnarray} \nonumber
\left\{
\begin{array}{l}
k_x(\theta)=  \cos^{-1} R(\mu,\theta)     \\
k_y(\theta)=     \sin^{-1} (- \mu \sin \theta )   \\
\end{array} \right.
\end{eqnarray}
for $\theta \in [-\sin^{-1} \frac{1}{\mu},0]$ and

 \begin{eqnarray} \nonumber
\left\{
\begin{array}{l}
k_x(\theta)= \pi-  \cos^{-1} R(\mu,\theta)   \\
k_y(\theta)= \pi+ \sin^{-1} (\mu \sin \theta )
\end{array} \right.
\end{eqnarray}
for $ \theta \in [\arccos (\frac{\mu^2+3}{4 \mu})-\pi ,\arcsin
(\frac{1}{ \mu})-\pi]$. The longitudinal variation $\Delta k_y$ is 
$\arccos \frac{\mu^2-5}{4} $ and the horizontal variation $\Delta k_x$ is $\arccos(\frac{\mu-1}{2})$. The resulting prefactors
of leading logarithmic terms are:
\begin{eqnarray}
&&c_{1,\textrm{zigzag}}(\mu)=\frac{1}{3\pi} \Delta k_x ,\nonumber \\
&&c_{1,\textrm{armchair}}(\mu)=\frac{1}{3\pi}  \Delta k_y .
\label{exactformofc1mu}
\end{eqnarray}
A generalization to anisotropic case and arbitrary edges is straightforward. The coefficient of the leading term in system of open boundary condition is half of the one of periodic boundary condition, due to the fact that the boundary between
the bipartite spaces is half of that of stripes \cite{Calabrese}.

\begin{figure}[!ht]
\begin{center}
 \includegraphics[width=7.5cm]{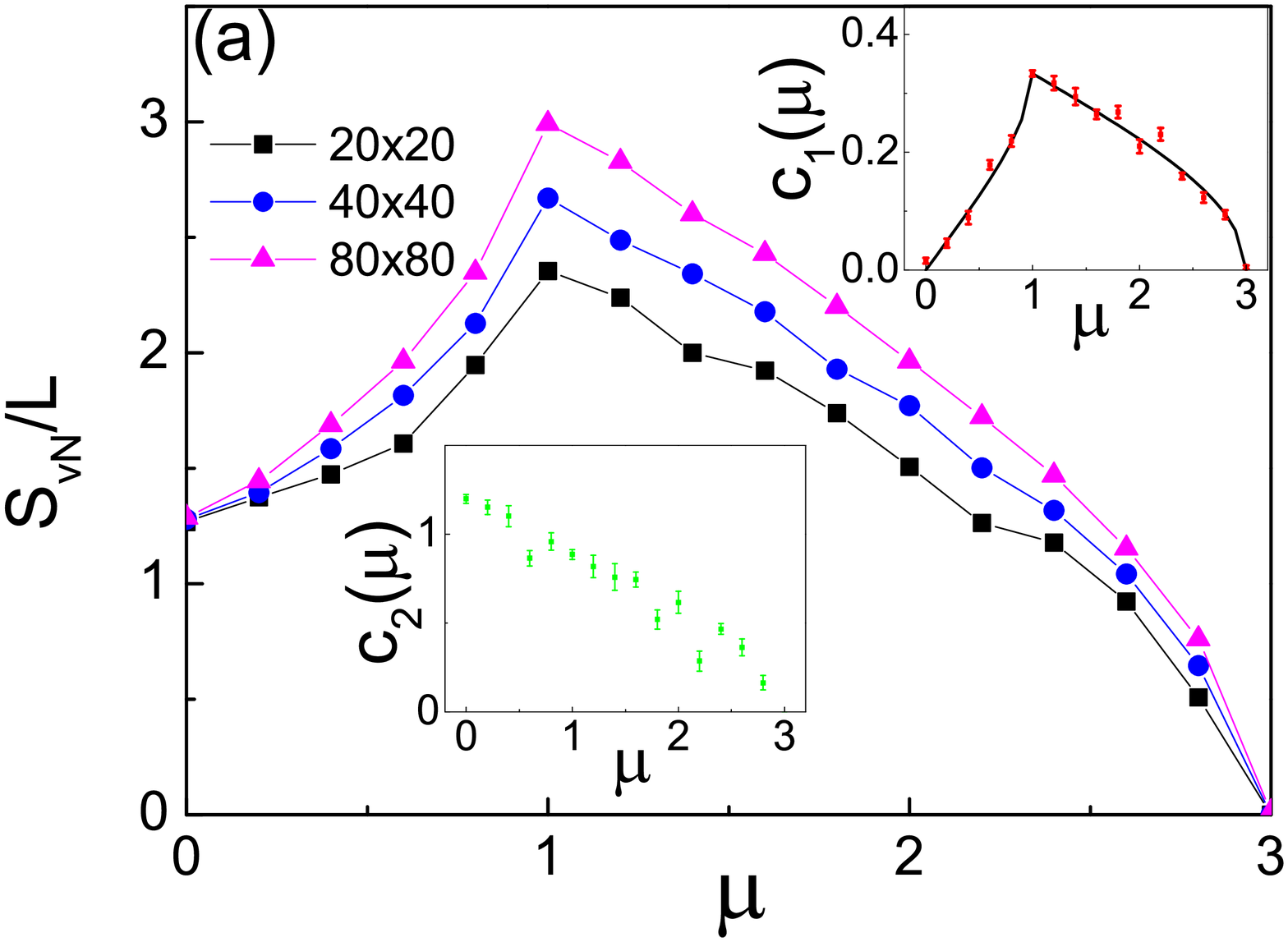}
 \includegraphics[width=7.5cm]{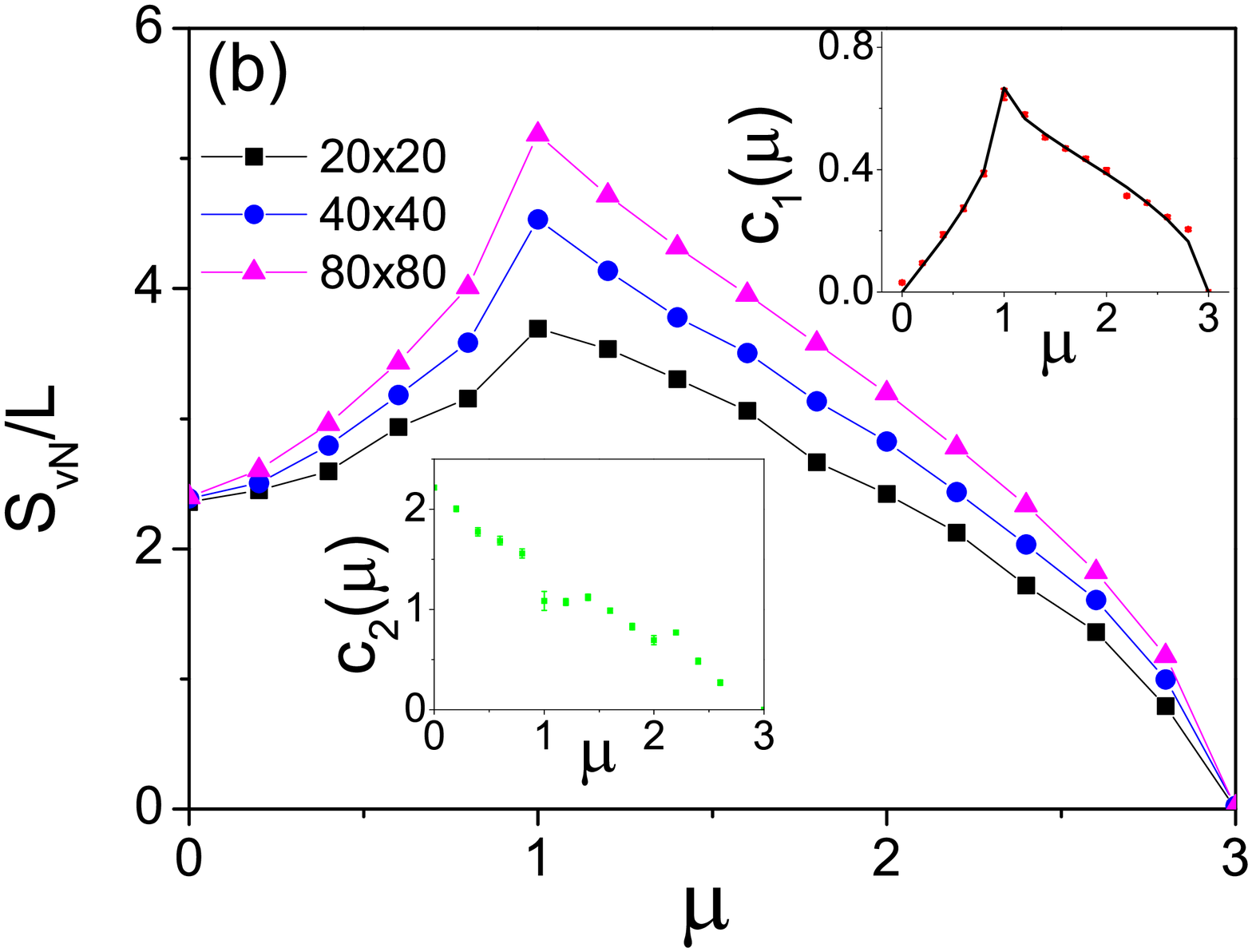}
\caption{(Color online) The entanglement entropy between $A$ and $B$
block for different chemical potential at $t_a$=$t_b$=$t_c$=1.0 with (a) zigzag edges and (b) armchair edges. The legends mark the half size. Insets show the prefactor $c_1(\mu)$, $c_2(\mu)$ in Eq.(\ref{logarithmicscaling})
as a function of the chemical
potential $\mu$ for the ground state of the two-dimensional
fermionic tight-binding model. $c_1(\mu$) extracted from fits to our numerical data (red filled square) are compared
with the analytical results (solid line) in the upper left inset. (c) Analytical results $c_1(\mu)$ for both edges, and the ratio in the inset.
}\label{SA-Ly320lA80-mu}
\end{center}
\end{figure}

From Fig. \ref{SA-Ly320lA80-mu}, we can recognize the entanglement entropy presents a shoulder-like behavior with respect to the chemical potential $\mu$. The finite-size effect becomes less dominated
for large size. $c_1(\mu)$ extracted from fits to our numerical data agrees rather well with the exact form Eq.(\ref{exactformofc1mu}) in the inset of Fig.\ref{SA-Ly320lA80-mu}. $c_1(\mu)$ goes to zero with $\mu$ approaches Dirac point, where the logarithmic correction is vanishing. When $\mu$ deviates slightly from the critical point, two small Fermi pockets are formed with Fermi vector $k_F=\vert \mu\vert/ v_F$, and Eq.(\ref{exactformofc1mu}) implies that $c_1(\mu)$ scales linearly for the low-energy modes when system size $L \gg l^*$, where $l^*$ is a characteristic length $l^*$ $\sim$ $k_F^{-1}$.  The ratios of $c_1(\mu)$ to $c_2(\mu)$  with both edges are above a factor $\sqrt{3}$, because the Brillouin zone of brick-type lattice along $y$ direction is stretched $\sqrt{3}$ times longer than that along $x$ direction comparing with pristine honeycomb lattice. The ratios reach their maxima passing through a Van Hove singularity, i.e., $\mu=1$, where two Fermi surfaces merge to form a single Fermi surface; see illustration in Fig.\ref{SA-Ly320lA80-mu}(c).
Thus at $\mu=1$ there is a Lifshitz transition. Besides a logarithmic divergence of the entanglement entropy, the linear-term coefficient is intricately related to ultraviolet cutoff or correlation length, and it is generally nonuniversal. Shown in the lower insets of Fig.\ref{SA-Ly320lA80-mu}  (a-b) are $c_2(\mu)$, which decrease with the increase of $\vert \mu \vert$. Extraordinarily, there is a dip at $\mu=1$ by a meticulous observation. We attribute such decrease to the enhanced contribution from a negative subarea term at $\mu=1$ \cite{Ding}.

\section{Spin-orbit interaction}
\begin{figure}[!ht]
\begin{center}
\includegraphics[width=7cm]{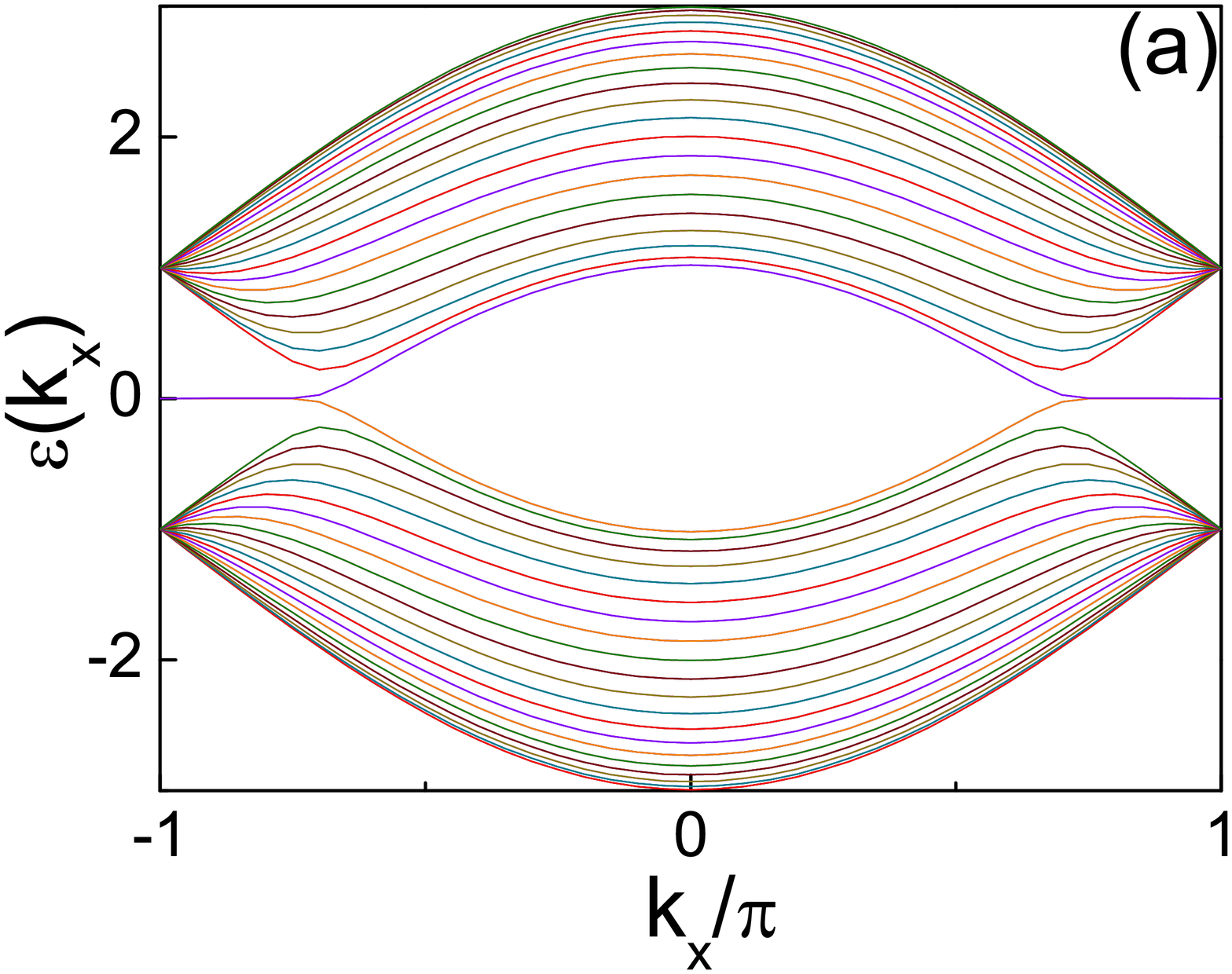}
\includegraphics[width=7cm]{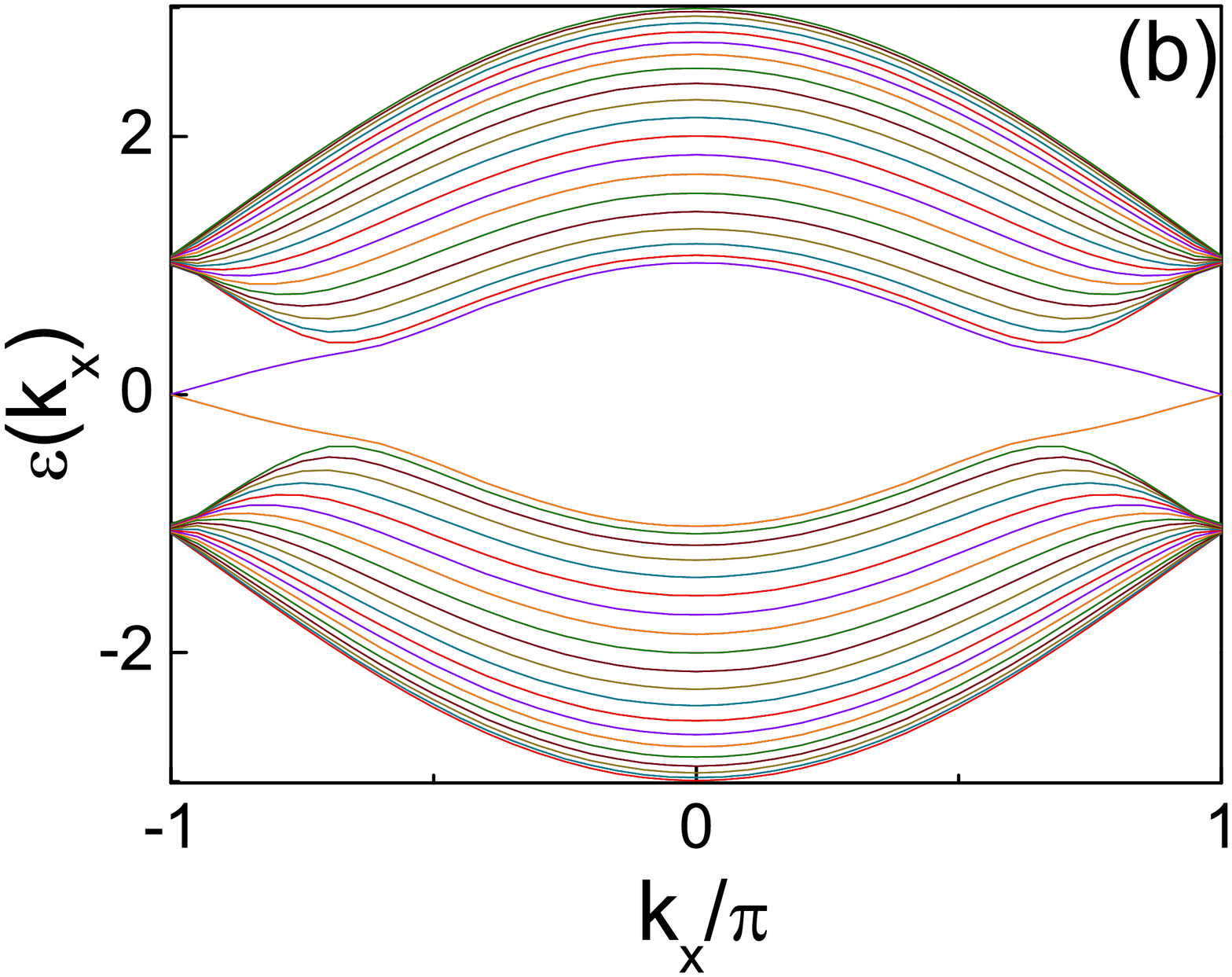}
\includegraphics[width=7cm]{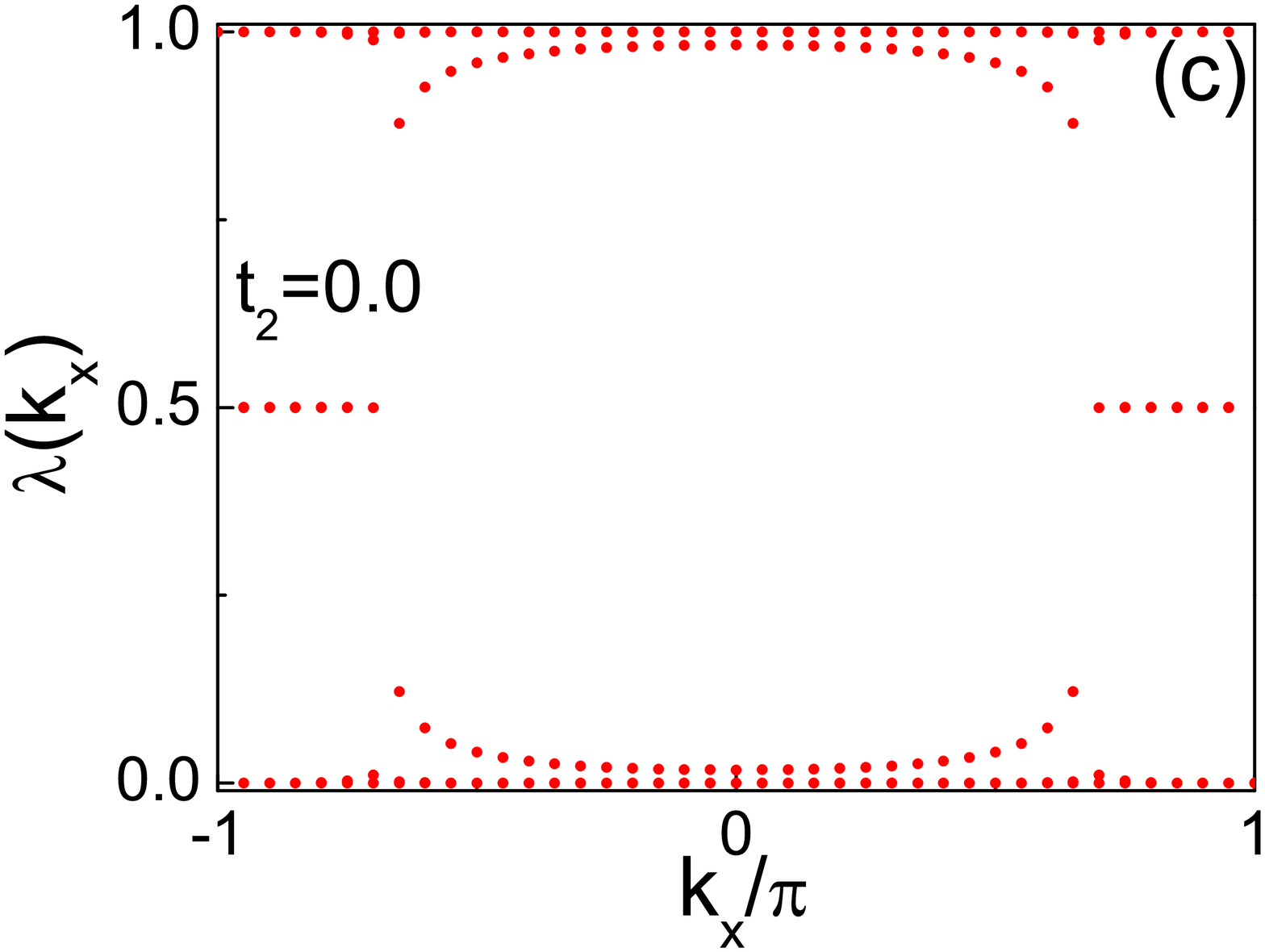}
\includegraphics[width=7cm]{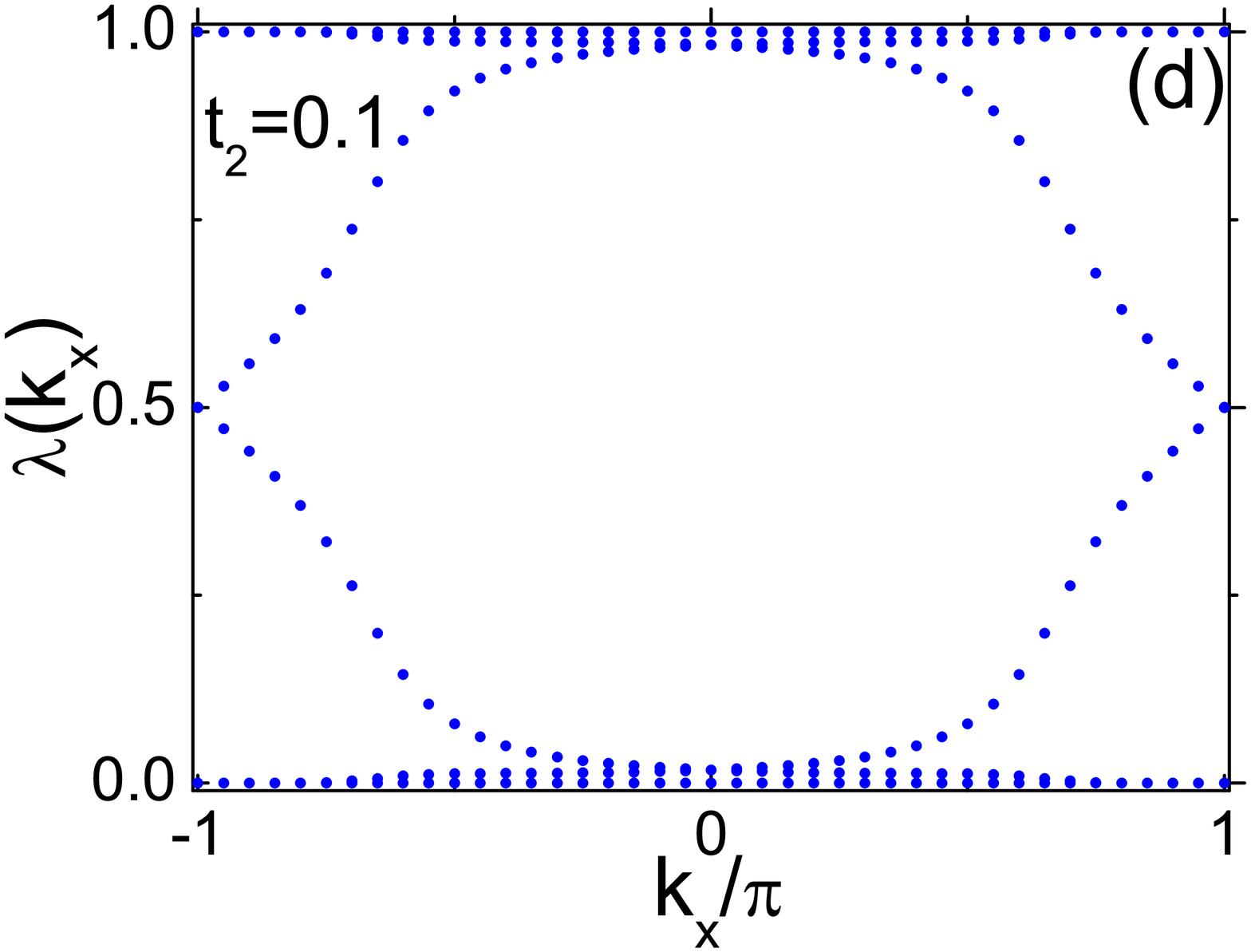}
\caption{(Color online) (Color online) The energy spectrum of Hamiltonian (\ref{SOgrapheneHamiltonian}) for (a) $t_2=0$ and (b) $t_2=0.1$ on $L_x=40$, $L_y=40$ lattice with zigzag edges. The corresponding entanglement spectrum at $\mu=0$ for (c) $t_2=0$ and (d) $t_2=0.1$. }\label{SOEES-Lx40Ly40}
\end{center}
\end{figure}

\begin{figure}[!ht]
\begin{center}
 \includegraphics[width=7cm]{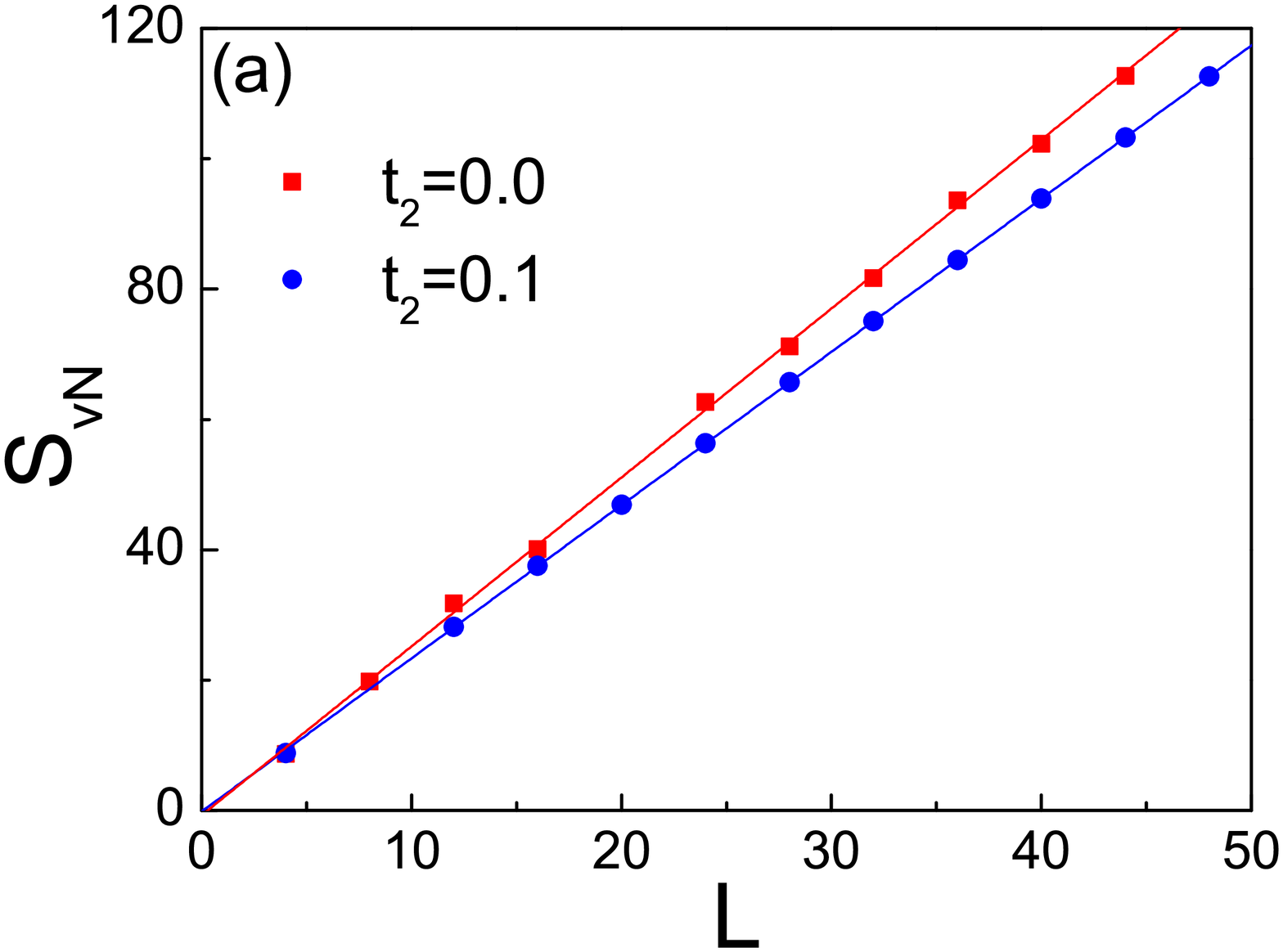}
  \includegraphics[width=7cm]{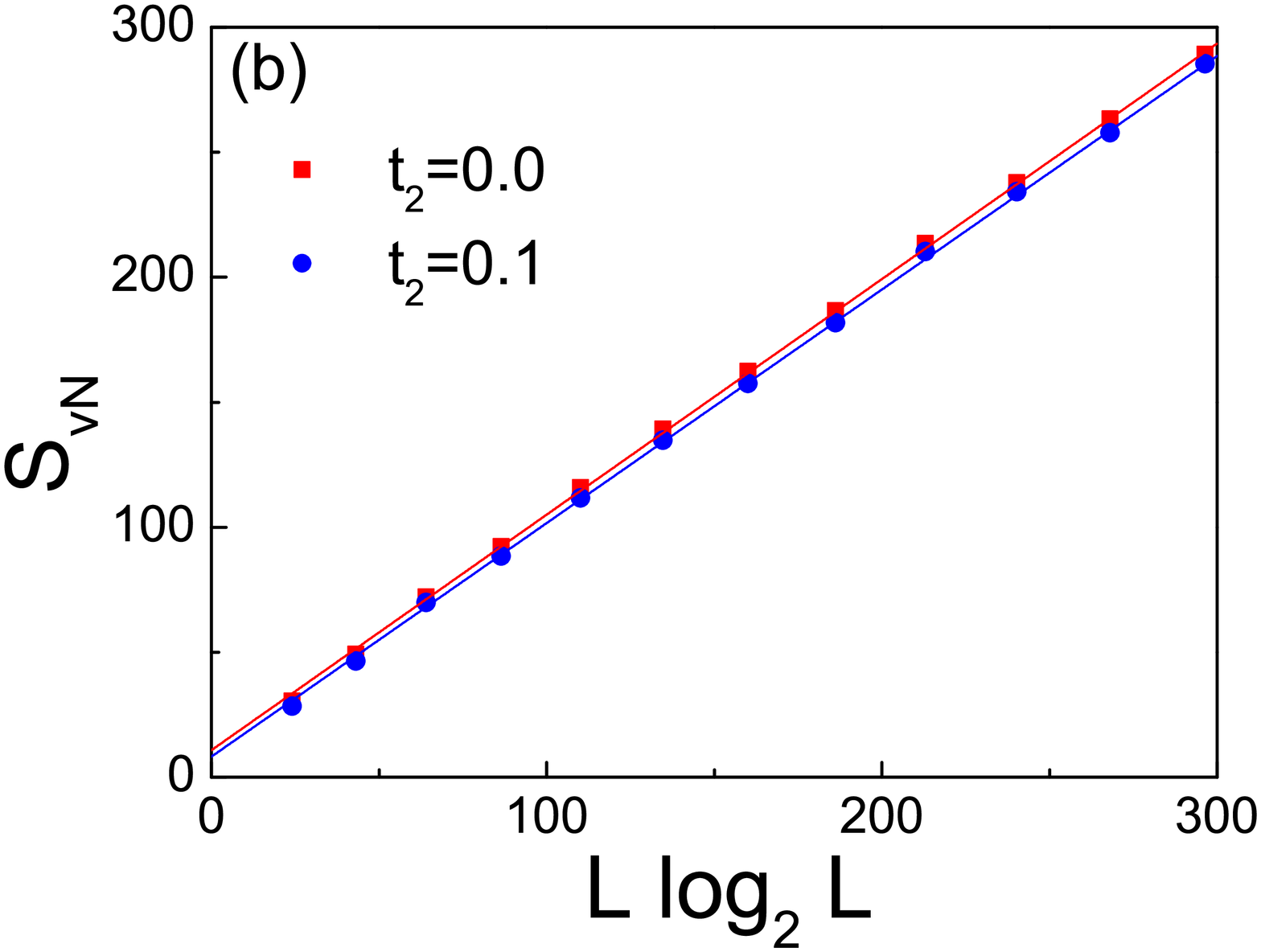}
\caption{(Color online) The scaling of entanglement entropy with zigzag edges at $t_2$=0 and 0.1 when (a) $\mu=0$ and (b) $\mu=-1$. Lines are linear fits.}\label{Scaling-EE-t1}
\end{center}
\end{figure}

Next, we extend our scope to Kane-Mele model, which is a tight-binding model with second neighbor spin-orbital interaction defined on a honeycomb lattice,
\begin{eqnarray}
H&=&-t \sum_{r\in \Lambda_A,\alpha}\sum_{i=a,b,c} (a_{r,\alpha}^{\dagger} b_{r+\tau_i,\alpha} +
b_{r+\tau_i,\alpha}^{\dagger} a_{r,\alpha} )\nonumber \\
&+& i t_2  \sum_{r\in \Lambda_A,\alpha,\beta}\sum_{i,j=a,b,c}  a_{r,\alpha}^{\dagger}[\boldsymbol{\sigma}
\cdot (\boldsymbol{\tau}_i \times \boldsymbol{\tilde{\tau}}_j)]_{\alpha,\beta} a_{r+\tau_i+\boldsymbol{\tilde{\tau}}_j,\beta}\nonumber \\
&-& i t_2  \sum_{r\in \Lambda_B,\alpha,\beta}\sum_{i,j=a,b,c}
b_{r,\alpha}^{\dagger} [\boldsymbol{\sigma}
\cdot (\boldsymbol{\tau}_i \times \boldsymbol{\tilde{\tau}}_j)]_{\alpha,\beta} b_{r+\tau_i+\boldsymbol{\tilde{\tau}}_j,\beta} .\nonumber\\
&-& \mu \sum_{r\in \Lambda_A,\alpha}a_{r,\alpha}^{\dagger} a_{r,\alpha}- \mu\sum_{r\in \Lambda_B,\alpha}b_{r,\alpha}^{\dagger} b_{r,\alpha}. \label{SOgrapheneHamiltonian}
\end{eqnarray}
Here $\tilde{\tau}_i=-\tau_i$ is the antiparallel vector of $\tau_i$ in Eq.(\ref{Bricksublatticeseparation}).
The Hamiltonian is time reversal invariant and has a wide range of
applicability, such as quantum spin-Hall
(QSH) insulator \cite{Kane1,Kane2}, graphene/graphane interface \cite{Schmidt}
and silicene \cite{Vogt,Fleurence,Lchen,Feng}.
The energy band structure of the brick lattice is given by
\begin{eqnarray}
\epsilon_k=\pm \sqrt{\Phi_k^2+\Theta_k^2},
\end{eqnarray}
where $\Phi_k = -t   (e^{i  k_y }+e^{i  k_x }+e^{-i  k_x })$ and
$\Theta_k = 4 t_2 \cos k_y  \sin k_x$.
Without the second-neighbor hopping, the model's bulk and surface
spectra display the non-generic feature that all energy
levels come in pairs at each momentum due to parity symmetry \cite{Turner}.
The presence of the spin-orbit
interaction, i.e., $t_2 \neq 0$, leads to a gap in the bulk spectrum, as shown in Fig.\ref{SOEES-Lx40Ly40}.
More interestingly, in the case of zigzag ribbon, the flat
dispersion is replaced by the helical modes intrinsic
to the QSH insulator as the spin-orbital is introduced \cite{Ezawa}.
The dominant contributions originates from
the helical edge state is displayed in Fig.\ref{SOEES-Lx40Ly40}(d).
The area-law monotonic scaling at $\mu=0$ is confirmed in Fig.\ref{Scaling-EE-t1}(a),
and the leading logarithmic divergence is also observed in Fig.\ref{Scaling-EE-t1}(b).

\section{Conclusion}

In this paper, we study the von Neumann entropy in free fermionic systems on a two-dimensional honeycomb lattice. Distinct bipartitions on the honeycomb lattice will produce different edges, and the edge states will have an effective impact upon the area-law terms in the entanglement entropy. We exhibit that the zero-energy edge state has an one-to-one correspondence with the maximal entangled mode in the entanglement spectrum, and hence we speculate the main contributions to the entanglement entropy come from the bulk states and edge states. The edge states play an essential role in semimetallic state with zigzag edges and in insulating phase with armchair edges. Furthermore, we find the entanglement entropy obeys area law when the Fermi surface is within the gap or a nodal point. The corrections to the pow-law term behave differently in critical phase from in noncritical phase. A logarithmic violation to the area law emerges in the presence of line-nodes Fermi surface. We also show that the logarithmic scaling highly depends on the topology of the Fermi surface. The prefactor of logarithmic term is determined by the variation of Fermi surface along the direction parallel the edge. The scaling laws are independent of the statistics of the microscopical
constituents. The introduction of spin-orbit coupling will cause a gap in the bulk spectrum and a decrease of entanglement entropy.

Since there is an equivalence between the entanglement entropy and particle number fluctuations in free-fermion systems \cite{Klich09,Song11,Song12}, it is very promising to
measure the entanglement entropy in the graphene-like systems. Especially experimentally the edges can be better fabricated in the artificial honeycomb than those of natural electronic graphene, which tend to be very irregular and contaminated with adsorbates. The flat edge modes can be observed in the zero-bias conductance measurements \cite{Hu}. Our results can be extended to other free fermionic systems, i.e., band topological insulators and superconductors, as well as weakly interacting systems \cite{Wenxin}.

\subsection*{Acknowledgments}
W-L.Y. thanks Peter Horsch and Andreas Schnyder for insightful discussions. This work was supported by the National Natural Science
Foundation of China (NSFC) under Grant No. 11004144.

\end{document}